\documentclass[12pt,a4paper,fleqn]{article}

\usepackage[DIV13]{typearea}
\usepackage[T1]{fontenc}
\usepackage{lmodern}
\usepackage{amsmath,amssymb}
\usepackage{graphicx}
\usepackage[footnotesize]{caption}
\usepackage{braket}
\usepackage[colorlinks,citecolor=blue]{hyperref}

\newcommand{\eV}{\ensuremath{\,\textnormal{eV}}}
\newcommand{\MeV}{\ensuremath{\,\textnormal{MeV}}}

\newcommand{\cm}{\ensuremath{\,\textnormal{cm}}}
\newcommand{\m}{\ensuremath{\,\textnormal{m}}}
\newcommand{\km}{\ensuremath{\,\textnormal{km}}}
\newcommand{\kpc}{\ensuremath{\,\textnormal{kpc}}}
\newcommand{\g}{\ensuremath{\,\textnormal{g}}}

\DeclareMathOperator{\sign}{sgn}

\newcommand{\Eqref}[1]{eq.~\eqref{#1}}
\newcommand{\Figref}[1]{fig.~\ref{#1}}
\newcommand{\Secref}[1]{sec.~\ref{#1}}

\newcommand{\be}{\begin{equation}}
\newcommand{\ee}{\end{equation}}

\allowdisplaybreaks[1]

\begin{document}

\begin{titlepage}

\vspace*{1cm}

\begin{center}
{\Large\bfseries
Decoherence and oscillations of supernova neutrinos
}

\vspace{1cm}
\renewcommand{\thefootnote}{\arabic{footnote}}

\textbf{
J\"orn Kersten\footnote[1]{Email: \texttt{joern.kersten@uib.no}}$^{(a)}$
and
Alexei Yu.\ Smirnov\footnote[2]{Email: \texttt{smirnov@ictp.it}}$^{(b,c)}$
}
\\[5mm]
\textit{\small $^{(a)}$
University of Bergen, Institute for Physics and Technology,\\
Postboks 7803, 5020 Bergen, Norway\\[2mm]
$^{(b)}$
Max-Planck-Institut f\"ur Kernphysik,\\
Saupfercheckweg 1, 69117 Heidelberg, Germany\\[2mm]
$^{(c)}$
The Abdus Salam ICTP, Strada Costiera 11, 34014 Trieste, Italy}
\end{center}

\vspace{1cm}

\begin{abstract}
\noindent
Supernova neutrinos have several exceptional 
features which can lead to interesting physical consequences. 
At the production point their wave packets have an extremely small size
$\sigma_x \sim 10^{-11}\cm$; hence the energy 
uncertainty can be as large as the energy itself, $\sigma_E \sim E$, and the coherence length is short. 
On the way to the Earth the wave  packets of mass eigenstates spread 
to macroscopic sizes and separate. Inside the Earth the mass eigenstates 
split into eigenstates in matter and oscillate again. 
The coherence length in the Earth is comparable with the
radius of the Earth. We explore these features and their consequences.
(i) We present new estimates of the wave packet size. 
(ii) We consider the decoherence condition 
for the case of wave packets with spatial spread and show that 
it is not modified by the spread.    
(iii) We study the coherence of neutrinos propagating in a multi-layer medium 
with density jumps at the borders of layers.
In this case coherence can be partially restored 
due to a ``catch-up effect'', increasing
the coherence length beyond the usual estimate.   
This catch-up effect can occur for supernova neutrinos as they cross the
shock wave fronts in the exploding star or the core of the Earth.
\end{abstract}

\end{titlepage}

%%%%%%%%%%%%%%%%%%%%%%%%%%%%%%%%%%%%%%%%%%%%%%%%%%%%%%%%%%%%%%%%%%%%
\newpage

\section{Introduction}

Detecting the neutrino burst from a galactic supernova will be one of
the major and out\-stand\-ing  scientific 
events of this century. 
It will bring an enormous amount of new physics information both on the dynamics of 
the collapse as well as the explosion and on neutrinos themselves. 
Hence, a deep understanding of the underlying 
processes and effects is a must. 

Supernovae -- the strongest known sources of neutrinos --  provide a
unique environment for the production  and 
the flavor evolution of neutrino states.   
In addition to the standard resonant flavor conversion 
in matter \cite{Dighe:1999bi},
the huge neutrino density makes neutrino--neutrino
interactions relevant \cite{Pantaleone:1992eq,Samuel:1993uw,Pantaleone:1994ns}
that can lead to various collective effects: 
synchronized oscillations \cite{Samuel:1993uw,Pastor:2001iu}, 
bipolar oscillations \cite{Duan:2005cp,Hannestad:2006nj},
spectral splits or swaps 
\cite{Duan:2006an,Raffelt:2007cb,Fogli:2007bk,Raffelt:2007xt},
self-induced parametric resonance \cite{Raffelt:2008hr},  etc.\
(see, e.g., \cite{Duan:2009cd,Duan:2010bg,Mirizzi:2015eza} for reviews).
Stimulated flavor transitions can occur due to turbulence 
in a medium \cite{Patton:2014lza}.

Given the special conditions with very high temperature and 
density in the neutrino\-sphere, where neutrinos are
produced, a very short time scale for the microscopic production
processes is realized.  Consequently, the neutrino states are
described by very short wave packets in configuration space.  Indeed,
previous estimates of the wave packet size were
$\sigma_x = 1.8 \cdot 10^{-14} \cm$ for neutrinos produced in the core
of the protoneutron star \cite{Anada:1987aw}
and $\sigma_x = 4.2 \cdot 10^{-9} \cm$ for neutrinos emitted at a
radius of $1000\km$ \cite{Anada:1989fk}
and thus in a region with much lower matter density.
These calculations used Coulomb scattering
to determine the mean free path of electrons inside the supernova.

Another important feature is the huge distance 
supernova neutrinos travel from a star to the Earth. 
Both these features, very short wave packets and very long baselines,
affect the flavor evolution of the neutrino states and consequently observations. 
In fact, two effects happen: 
\begin{itemize}
\item 
shift of the wave packets
of the eigenstates by $\Delta x_\text{shift}$ due to the
difference of group velocities and eventually their separation  
when $\Delta x_\text{shift} > \sigma_x$, implying a loss
of coherence of the neutrino states, and
\item
spread of wave packets of individual eigenstates 
due to the presence of different energies in a wave packet. 
\end{itemize}
Due to the very small $\sigma_x$, decoherence may occur at small distances
inside the supernova, even before the region of collective effects.
Due to the huge distance to the Earth, the separation and  
spread of the packets become macroscopic. So, in principle, one can 
discuss the possibility of time tagging the wave packets.

Propagation decoherence has been studied studied mainly in vacuum
\cite{Kayser:1981ye,Kiers:1995zj,Grimus:1998uh,Cardall:1999ze,Giunti:2000kw,Ohlsson:2000mj,Giunti:2003ax,Akhmedov:2009rb,Akhmedov:2012uu},
but also in matter
\cite{Mikheyev:1989dy,Anada:1989fk,Peltoniemi:2000nw,Hollenberg:2011tc,Oliveira:2014jsa}
and in dense neutrino gases \cite{Raffelt:2010za}.
It is characterized by the coherence 
length $L_\text{coh}$ -- the distance at which 
$\Delta x_\text{shift} \sim \sigma_x$. 
For distances larger than $L_\text{coh}$, the eigenstates no longer
interfere because their wave packets no longer overlap.  Thus, the
oscillatory 
pattern disappears and the oscillation probability becomes
baseline-independent.

The coherence is also affected by the detection process.
In particular, coherence can be restored  
by an accurate energy measurement \cite{Kiers:1995zj},
which was confirmed  via a quantum field theory calculation
\cite{Giunti:1997sk}.  In this case the detector must have a
coherent observation time larger than the difference of the arrival
times of two packets $\Delta t_\text{det}$ or (equivalently) a
sufficiently good energy resolution,
$\Delta E_\text{res} < 1/\Delta t_\text{det}$. 
Thus, observable effects depend on characteristics of the detector:
coherent time of observation and energy resolution.

The wave packets give the complete picture of the evolution of neutrino states in
configuration space. However, in order to determine observational results
it is enough to treat the problem in the energy representation, which 
simplifies the considerations substantially.
The results of the energy-momentum and configuration space 
treatments are equivalenct as far as observations
are concerned, at least in the stationary source approximation
\cite{Kiers:1995zj,Stodolsky:1998tc} 
and if no time tagging is performed. 

The effects of source and detector are symmetric and can be described by an effective
wave packet which includes characteristics of both source and detector. 
In fact, information about the wave packet of the source is included in
the generated energy spectrum and information about the one of the
detector in the detector's energy resolution.

It is believed that propagation decoherence does not affect the dynamics 
of the flavor evolution.  Observable effects are then determined
essentially by the energy spectrum at production, flavor evolution
without decoherence, and the energy resolution of the detector.  
As mentioned above, however, the separation of supernova neutrinos is
very fast due to their very short wave packet size.  This could affect
the dynamics of the flavor evolution in the region of collective effects
\cite{Kersten:2013fba}, although a tentative answer was negative. In
contrast, in \cite{Akhmedov:2014ssa} it was claimed that decoherence 
does influence collective oscillations 
(essentially due to the non-linear character of the problem)  
and lead to non-trivial flavor transformations.   
 
Propagation decoherence is a reversible process; no information 
is lost in a system when wave packets separate.
Hence, coherence can be restored by further propagation in matter if
the difference of group velocities changes
sign or (as mentioned above) in the detector.

The consideration in configuration space may have some advantage in the
case of a complicated matter profile.  It helps to obtain a clear
physics interpretation of the results of integrating over energy.  We
will consider in detail the evolution of the wave packets all the way
from the production point to a detector, focussing on the consequences
of the small wave packet size of supernova neutrinos.  We will study
decoherence and partial restoration of coherence.  New interesting
effects are realized in oscillations in the matter of the Earth, which
are related to an accidental coincidence of the coherence length and the
size of the Earth.  Some preliminary results have been published in
\cite{Kersten:2013fba}.

Apart from separating, the wave packets also spread, since they comprise
waves with different energies~\cite{Beuthe:2001rc,Chan:2015mca}.  The
increase of the size of a packet depends on the absolute values of
neutrino masses.  Effects of the spread on oscillations (which were not
explored extensively before) are among the main objectives of this work.

The paper is organized as follows. In \Secref{sec:CoherenceLength} we 
consider the production of neutrinos in a supernova and present an
improved estimate of their wave packet size.  We consider propagation
decoherence, generalizing the vacuum results to propagation in matter.
In \Secref{sec:Averaging} we consider decoherence in the energy-momentum
representation.  We show the equivalence of separation of the packets in
configuration space and energy averaging. 
In \Secref{sec:Cohandaddv} we study flavor evolution and coherence in a
multi-layer medium where neutrino states split at each border between
layers. We describe the ``catch-up effect'' -- partial restoration
of the coherence between certain components of the split states.
In \Secref{sec:Spread} we discuss the spread of wave packets both in
vacuum and in matter.  We show that the coherence conditions for wave
packets with and without spread coincide. 
In \Secref{sec:EarthOsci} we apply our results to supernova neutrino
oscillations in the matter of the Earth.

\section{Supernova neutrino wave packets}
\label{sec:CoherenceLength}
\subsection{Size of the wave packet}
%%%%%%%%%%%%%%%%%%%%%%%%%%%%%%%%%%%%%%%%%%%%%%%%%%%%%%%%%%%%%%%%%%%%%%%%%

The characteristics of the neutrino wave packets (WP) 
produced in a supernova depend on the phase of the explosion,
since physical conditions and contributing processes change. 
We will consider different phases in order. 

\paragraph{Neutronization burst.}
During the earliest stage of a supernova, mainly electron neutrinos
are produced by electron capture, $p~ e^- \to n ~\nu_e$.
We assume the nucleons to be localized well enough that the size of
their WP is negligible (see below).  Then the time scale for
the electron capture process is given by the interval of time during
which the electron WP crosses the proton \cite{Kiers:1995zj}: 
$\tau \simeq \sigma_x^e / v^e \simeq \sigma_x^e$ for
relativistic electrons. Here $\sigma_x^e$ is the size of the electron
WP in configuration space and $v_e$ is the electron velocity.
During this time a neutrino is emitted coherently.  Consequently,
its WP has a size  $\sigma_x \simeq \tau \simeq \sigma_x^e$.%
\footnote{We define the WP width as the position or energy uncertainty
 of a particle.
 We employ the ``intermediate wave packet'' picture for the neutrinos.
 This approach produces the correct results for oscillation
 probabilities, as shown by the quantum field theory treatment of
 neutrino oscillations in vacuum
 \cite{Giunti:1997sk,Beuthe:2001rc,Giunti:2002xg}.
 We consider ultra-relativistic neutrinos throughout.
\label{fn:WP}}

In turn, we estimate the electron WP size as $\sigma_x^e \simeq \lambda_e$, where
$\lambda_e$ is the mean free path between two collisions of an electron that change
its momentum by more than the momentum uncertainty 
$\sigma_p^e \simeq 1/\sigma_x^e$.  
Collisions with a smaller momentum transfer correspond to forward
scattering, which does not modify the WP and thus cannot localize a
state. 
Introducing the electron fraction $Y_e$ 
we have $n_e \simeq Y_e n_N$, where $n_N$ is the nucleon number
density, which is related to the mass density $\rho$ by
$n_N \simeq N_A \rho / (1\g\cdot\text{mol}^{-1})$, where
$N_A$ is the Avogadro constant.  We neglect the Pauli blocking and consider
only electron--electron as well as electron--proton scattering.
We expect electron--positron and Compton scattering to be subdominant
due to the significantly smaller densities of the respective scattering
partners.  Approximating the cross section for $ee$ scattering by the
result in the center-of-mass frame and the one for $ep$ scattering by
the result for infinitely heavy protons we obtain 
\begin{equation} \label{eq:eeScattering}
	\sigma(\Delta p > \sigma_p^e) \simeq
	\frac{4\pi\alpha^2}{{\sigma_p^e}^2}
\end{equation}
for both processes.  Then
\begin{equation} \label{eq:MFP}
	\frac{1}{\lambda_e} \simeq 
	\frac{8\pi\alpha^2}{{\sigma_p^e}^2} \cdot Y_e n_N \simeq
	8\pi\alpha^2 {\sigma_x^e}^2 \, Y_e n_N \;,
\end{equation}
since the electron and proton number densities are roughly equal.
Using $\sigma_x^e \simeq \lambda_e$ 
we obtain the total width of the neutrino WP $\sigma_x \simeq \sigma_x^e$:
\begin{equation} \label{sigmax}
\sigma_x \simeq \left( 8\pi \alpha^2 Y_e n_N \right)^{-1/3} \simeq
1.4 \cdot 10^{-11} \cm
\left( \frac{10^{12} \g/\text{cm}^3}{2 Y_e \, \rho} \right)^{1/3} .
\end{equation}
Note that the third root of the density and cross-section enters this
expression, so the order of magnitude of the result should be correct
despite our rather crude approximations.  For a density 
$\rho = (10^{11} - 10^{12}) \g / \text{cm}^3$
and an electron fraction $Y_e = 0.5$ in the production region
(neutrinosphere) we obtain 
\begin{equation} \label{eq:DeltaxSN}
	\sigma_x \simeq (1.4 - 3) \cdot 10^{-11} \cm \;,
\end{equation}
or in energy-momentum space
\begin{equation} \label{eq:DeltaESN}
	\sigma_E \simeq \frac{1}{\sigma_x} \simeq (0.7 - 1.5) \MeV \;.
\end{equation}
Notice that these values do not depend on the neutrino energy. 
Incidentally, the thermal wavelength, which has been suggested as an
alternative measure of the neutrino position uncertainty
\cite{Kiers:1995zj}, yields a  similar value,
\begin{equation} \label{eq:ThermalWavelength}
	\sigma_x \sim \lambda_T = \frac{2\pi}{\braket{p}} \sim
	\frac{2\pi}{3\,T} \sim 8 \cdot 10^{-12}\cm
\end{equation}
for $T\sim5\MeV$. 

For nucleons,  which are non-relativistic, we estimate the size of WP from the thermal wavelength
\begin{equation}
	\sigma_x^N \sim \lambda_T^N \sim
	\frac{2\pi}{\sqrt{3m_NT}} \sim 10^{-12}\cm \;,
\label{eq:nuclx}
\end{equation}
which is of the same order of magnitude as the mean distance
between nucleons $d \sim n_N^{-1/3} \sim (1 - 3) \cdot 10^{-12} \cm$ 
and much smaller than the width of the electron WP\@.
Hence, the nucleon WP size does not play a role in our discussion,
justifying our initial assumption.
Note that the magnetic fields inside a supernova are too weak to change
the momentum of an electron between two collisions significantly, so
they do not influence the WP size.

\paragraph{Accretion and cooling phase.}
For $\nu_e$,
electron capture remains the main production process.  Consequently, the
WP size is still given by \Eqref{sigmax}.  Compared to the
neutronization burst, the physical parameters in the neutrinosphere
change in opposite directions:  while the matter density $\rho$
increases, the electron fraction $Y_e$ decreases by about an order of
magnitude \cite{Tamborra:2011is,Fischer:2011cy}.  As a result, we expect
the $\nu_e$ WP size to increase moderately.

Electron antineutrinos $\bar\nu_e$ are mainly produced by positron
capture on neutrons, $n ~ e^+ \to p ~ \bar\nu_e$, so their $\sigma_x$ is
equal to the WP size of the positron.  The size of the positron~WP can
be estimated in the same way as that of the electron WP\@.  The most
important scattering processes for positrons are scatterings on
electrons and on protons, whose cross section is the same as the one for
electron ($ee$ and $ep$) scattering, \Eqref{eq:eeScattering}.  As a
consequence, \Eqref{sigmax} holds for $\bar\nu_e$ as well.

For the non-electron neutrinos, $\nu_\mu$ and $\nu_\tau$, and their
antineutrinos, the physics of production is quite different.  The number
density of these neutrinos is determined at the number sphere.  Outside
this sphere neutrinos still scatter efficiently on leptons (until the
energy sphere) and on nucleons (until the transport sphere)
\cite{Keil:2002in}.  Hence, the WP size of non-electron neutrinos is
determined by these scattering processes.  In general, the size of the
WP after a scattering process is determined by the time of overlap of
the incoming WP and thus approximately equal to the size of the
\emph{larger} incoming WP \cite{Beuthe:2001rc}.  As a consequence,
neutrino WP will continue to broaden as long as they scatter with
particles with larger WP sizes.  In our case these particles are the
leptons, since the nucleon WP are estimated to be very small, see
\Eqref{eq:nuclx}.  Consequently, we expect $\sigma_x$ for $\nu_\mu$ and
$\nu_\tau$ to equal the electron WP size at the energy sphere.
This means that once again the result is given by \Eqref{sigmax}.

\subsection{Propagation decoherence} \label{sec:PropDec}
%%%%%%%%%%%%%%%%%%%%%%%%%%%%%%%%%%%%%%%%%%%%%%%%%%%%%%%%%%%%%%%%

Propagation of (flavor) mixed states is described by a system of the WP
which correspond to the eigenstates of propagation, i.e.,
the eigenstates of the Hamiltonian in matter,
$\nu_{im}$ ($i=1,2,3$).%
\footnote{The eigenstates of the Hamiltonian should be considered the ``true
 particles'', in analogy to the concept of quasi-particles in
 condensed matter physics \cite{Cardall:2001cw}.
}
The group velocities of the eigenstates  
are determined by the eigenvalues of the Hamiltonian: 
\begin{equation} 
\label{eq:GroupVelocity}
v_{im} = \frac{dH_{im}}{dp} \;.
\end{equation}

We will consider a two-neutrino system characterized by a mass squared
difference $\Delta m^2 \equiv m_2^2 - m_1^2$ and a vacuum mixing angle $\theta$.
Using the explicit expressions for $H_{im}$ in the two-neutrino case, we find 
the difference of the eigenvalues
\begin{equation} 
\label{eq:DeltaHm}
	\Delta H_m  \equiv H_{2m} - H_{1m} = \frac{\Delta m^2}{2E} R(\xi) \;,
\end{equation}
where
\be
R(\xi) \equiv \sqrt{(\cos 2\theta - \xi )^2 + \sin^2 2\theta}
\label{eq:resfactor}
\ee
and\footnote{We use the variable $\xi$ instead of $\eta = 1/\xi$ in
\cite{Mikheyev:1989dy}.}
\begin{equation}
	\xi \equiv \frac{2EV}{\Delta m^2}
\end{equation}
with the matter potential $V = \pm\sqrt{2} \, G_F n_e$ for neutrinos and
antineutrinos, respectively.  Here $G_F$ is the Fermi coupling constant.
In vacuum $R = 1$. 

Due to the difference of the group velocities,
\be
\Delta v_m \equiv v_{1m} - v_{2m} = -\frac{d \Delta H_m}{dp} \;,
\label{eq:diffvel}
\ee
the WP shift with respect to each other in configuration space.  
After travelling a distance~$L$ the shift equals 
\begin{equation} 
\label{eq:delx}
	\Delta x_\text{shift} = \int_0^L dx \, \Delta v_m \;,
\end{equation}
which is $\Delta v_m L$ for a medium with constant density.
If the size of the WP does not change in the course of propagation, they
cease to overlap when $|\Delta x_\text{shift}| \simeq \sigma_x$,
which is called propagation decoherence.  The distance $L_\text{coh}$ at
which this happens, the coherence length~\cite{Nussinov:1976uw},
is determined by 
\begin{equation} \label{eq:LCohPosSpace}
	\left| \int_0^{L_\text{coh}} dx \, \Delta v_{m} \right| = \sigma_x \;.
\end{equation}
For constant density this gives 
\be 
	L_\text{coh} = \frac{\sigma_x}{|\Delta v_m|} \;.
\label{eq:cohlength}
\ee

In vacuum eq.~(\ref{eq:delx}--\ref{eq:cohlength}) 
are reduced to well-known results \cite{Kayser:1981ye}:
the eigenstates of the Hamiltonian are the mass eigenstates, $\nu_{im} = \nu_i$, 
the difference of group velocities equals
\begin{equation} \label{eq:DeltavmEarth}
\Delta v = \frac{\Delta m^2}{2 E^2} \;,
\end{equation}
and the coherence length according to \Eqref{eq:cohlength} is
\begin{equation} \label{eq:lcohvac}
L_\text{coh} = \sigma_x \frac{2E^2}{\Delta m^2} \;.
\end{equation}
In the three-neutrino system there are three different coherent lengths which correspond to 
three difference modes of oscillations driven by three different
$\Delta m^2_{ij} \equiv m^2_i - m^2_j$.

In matter with constant or slowly changing density,
the difference of velocities of the eigenstates equals~\cite{Mikheyev:1989dy}
according to (\ref{eq:diffvel}) and (\ref{eq:DeltaHm}) 
\begin{equation} 
\label{eq:Deltav}
	\Delta v_{m} = \frac{\Delta m^2}{2E^2} \,
	 \frac{1 - \xi \cos2\theta}{\sqrt{1 -2\xi\,\cos2\theta + \xi^2}} = 
	\frac{\Delta m^2}{2E^2} \, \frac{1 - \xi \cos2\theta}{R(\xi)} \;. 
\end{equation}
Equation (\ref{eq:Deltav}) gives $\Delta v_m \simeq \Delta v$ 
in the limit of small densities, $|\xi| \ll 1$, and   
\begin{equation} \label{eq:DeltavmHighDensity}
	\Delta v_m \simeq -\sign(\xi) \, \frac{\Delta m^2}{2E^2} \cos2\theta =
	-\sign(\xi) \, \Delta v \cos2\theta  
\end{equation}
in the matter-dominated case, $|\xi| \gg 1$.  For small mixing we obtain
again the vacuum value, up to a possible sign change.  This is related
to the fact that the matter potential $V$ does not depend on energy and
therefore does not produce dispersion.  Correspondingly, $L_\text{coh}$
is close to the vacuum value everywhere apart from the resonance region
in the resonance channel with $\xi > 0$. This is realized 
for neutrinos and the  normal mass
hierarchy, and for antineutrinos and the inverted mass hierarchy.  
In the non-resonance channel, $\xi < 0$, (i.e., for antineutrinos and a normal mass
hierarchy,  and for neutrinos and  the inverted hierarchy), the 
difference of  velocities 
and coherence length are close to the vacuum values everywhere.

In matter with  density such that 
\be
\xi =  \xi_0 \equiv \frac{1}{\cos2\theta} \;,
\ee
$\Delta v_{m} = 0$ \cite{deHolanda:2003nj}, so 
the WP do not shift and separate. 
In this case $L_\text{coh} \to \infty$.%
\footnote{The WP still spread, see \Secref{sec:Spread}.}
We will call $\xi_0$ the critical value and the corresponding density
and energy the critical density and critical energy.  The critical value
is larger than the resonance value $\xi_R = \cos2\theta$, as
$\xi_0 = \xi_R \,/ \cos^2 2\theta$.  For small mixing it is close to
$\xi_R$.

When neutrinos propagate from
large to small densities, $\Delta v_m$ changes sign near the MSW resonance.
This can lead to the interesting phenomenon that WP separate above (in density)
the resonance, then approach each other below the resonance and overlap at some point, 
thus restoring coherence~\cite{Mikheyev:1989dy,deHolanda:2003nj}.
This can be realized  for solar and  supernova neutrinos.   

In the case of quickly changing density (strong adiabaticity violation) 
the instantaneous eigenstates may become irrelevant for the description of 
the flavor evolution.  So the WP and group velocities introduced 
for these eigenstates have limited (or no) sense.

\subsection{Wave packet separation and coherence loss in a supernova}
%%%%%%%%%%%%%%%%%%%%%%%%%%%%%%%%%%%%%%%%%%%%%%%%%%%%%%%%%%%%%%%%%%%%

When propagating from the neutrinosphere to the surface of the star,
neutrinos cross regions with changing  conditions which affect
propagation and coherence of WP\@.  In the central parts,
neutrino--neutrino scattering leads to the potential $V_{\nu\nu}$
\cite{Pantaleone:1992eq,Pantaleone:1994ns}.  This potential is much
smaller than the usual matter potential $V$ in the neutrinosphere, but
it can be comparable to $V$ or even bigger at distances of order
$10^2\km$ from the center during later phases of the supernova explosion.  
$V_{\nu\nu}$ depends on the neutrino flavor state (i.e., on the neutrino wave 
function),
which leads to the so-called collective oscillations.

Outside the regions and time period where $V_{\nu\nu}$ is important we
can use the results of the previous subsection.
Above resonances  the difference of group velocities is given 
by \Eqref{eq:DeltavmHighDensity} and does not depend on density. 
So,  the integration in \Eqref{eq:LCohPosSpace} is trivial, and
consequently  we obtain for the coherence length
\begin{equation} \label{eq:lcohvac1}
	L_\text{coh} \simeq \sigma_x \frac{2E^2}{\Delta m^2 \cos2\theta} \;.
\end{equation}
Using the range for $\sigma_x$ from \Eqref{eq:DeltaxSN},
this gives for the 1-3 oscillation mode driven by 
$|\Delta m^2_{31}| \simeq 2.5 \cdot 10^{-3} \eV^2$ and
$\sin^2\theta_{13} \simeq 0.022$
as well as for the 1-2 mode with
$\Delta m^2_{21} \simeq 7.5 \cdot 10^{-5} \eV^2$ and
$\sin^2\theta_{12} \simeq 0.30$ \cite{Gonzalez-Garcia:2014bfa}
\begin{align}
	L_\text{coh}^{(13)} &\simeq
	(30 - 60) \km \left( \frac{E}{15\MeV} \right)^2 ,
	\label{eq:LCoh13}
\\
	L_\text{coh}^{(12)} &\simeq
	(2.1 - 4.5) \cdot 10^3 \km \left( \frac{E}{15\MeV} \right)^2 .
	\label{eq:LCoh12}
\end{align}
The estimate \eqref{eq:LCoh13} shows that the coherence length
$L_\text{coh}^{(13)}$ of supernova neutrinos is shorter than or
similar to the distance to the region where flavor evolution starts,
in particular where collective effects due to $\nu\nu$ scattering
become operative. This motivates studying decoherence effects on
collective oscillations.

%%%%%%%%%%%%%%%%%%%%%%%%%%%%%%%%%%%%%%%%%%%%%%%%%%%%%%%%%%%%%%%%%%%%%%%
\section{Decoherence and averaging over neutrino energy}
\label{sec:Averaging}
%%%%%%%%%%%%%%%%%%%%%%%%%%%%%%%%%%%%%%%%%%%%%%%%%%%%%%%%%%%%%%%%%%%%%%%

So far we have considered WP in configuration space.  We have also
assumed that the coherence length is determined entirely by the
production process.  Alternatively, we can consider decoherence in
energy-momentum space, where the WP width is
$\sigma_E \simeq \sigma_p \simeq 1/\sigma_x$.
This consideration makes it easier to take into account the detection process.

\subsection{Coherence in energy-momentum space} 
\label{sec:LCohMomSpace}
%%%%%%%%%%%%%%%%%%%%%%%%%%%%%%%%%%%%%%%%%%%%%%%%%%%%%%%%%%%%%%%%%%%

We generalize here  known results for propagation in vacuum.  
The oscillation phase is
\begin{equation} 
\label{eq:OsciPhase}
	\phi(E,L) = \frac{\Delta m^2 L}{2E} \;, 
\end{equation}
and the interference term in the oscillation probability is proportional
to $\cos\phi$.  This term leads to the characteristic oscillatory
pattern that can be observed in experiments.
The probability has to be averaged over the size of the
WP in momentum space.  This averaging suppresses the interference term
if the spread of phases within the WP is larger than $2\pi$.
Consequently, the coherence length can be defined via the condition 
\begin{equation} 
\label{eq:LCohMomSpace}
	|\Delta \phi| =
	\left|
	 \phi(E-\sigma_E,L_\text{coh}) - \phi(E+\sigma_E,L_\text{coh})
	\right| = 2\pi \;.
\end{equation}
Using \Eqref{eq:OsciPhase} we obtain  
\begin{equation}
	L_\text{coh} =
	\frac{2\pi}{\sigma_E} \frac{E^2}{|\Delta m^2|}
	 \left(1 - \frac{\sigma_E^2}{E^2} \right) \simeq
	\frac{\pi}{\sigma_E} \frac{2E^2}{|\Delta m^2|} \;.
\end{equation}
This result for the coherence length has the same form as
\Eqref{eq:lcohvac} but is larger by a factor of $\pi$.
Given the fact that the definitions \eqref{eq:LCohPosSpace} and
\eqref{eq:LCohMomSpace} are ad hoc and do not take into account that
coherence is not lost abruptly, it is not surprising that the results
for $L_\text{coh}$ agree up to a factor of order
one.  In the rigorous quantum field theory treatment, no such
discrepancy arises \cite{Giunti:1997sk,Beuthe:2001rc,Giunti:2002xg}.

We can immediately generalize the discussion to the case of 
matter with adiabatically varying density.  The oscillation phase is now
\begin{equation}
	\phi(E,L) = \int_0^L dx \, \Delta H_m \;,
\end{equation}
where $\Delta H_m$ is given in \Eqref{eq:DeltaHm}.  Taylor-expanding
$\phi$ in \Eqref{eq:LCohMomSpace}, which is
justified as long as $\sigma_E/E \ll 1$, 
we obtain 
\begin{equation} 
\label{eq:LCohMomSpaceVar}
	\Delta\phi \simeq
	-2\sigma_E \left.\frac{\partial\phi}{\partial E}\right|_{E,L_\text{coh}} =
	-2\sigma_E \, \frac{\partial}{\partial E} \int_0^{L_\text{coh}} dx \, \Delta H_m =
	-2\sigma_E \int_0^{L_\text{coh}} dx \, \frac{d}{dE} \Delta H_m \;.
\end{equation}
Then the condition $|\Delta\phi|=2\pi$ gives 
\begin{equation} 
\label{eq:cohener}
	\left| \int_0^{L_\text{coh}} dx \, \frac{d}{dE} \Delta H_m \right| \simeq
	\frac{\pi}{\sigma_E} \;.
\end{equation}

On the other hand, according to \Eqref{eq:GroupVelocity} and
\eqref{eq:delx}, the separation of the two WP in configuration space is%
\footnote{As neutrinos are highly relativistic, $d/dE \simeq d/dp$.}
\begin{equation}
	\Delta x_\text{shift} =
	-\int_0^L dx \, \frac{d}{dE} \Delta H_m \;.
\label{eq:sepgen}
\end{equation}
Therefore the decoherence condition $|\Delta x_\text{shift}| = \sigma_x$ 
yields
\begin{equation}
	\left| \int_0^{L_\text{coh}} dx \, \frac{d}{dE} \Delta H_m \right| =
	\sigma_x \simeq \frac{1}{\sigma_E} \;,
\end{equation}
which is the same condition as \Eqref{eq:cohener},
again up to a spurious factor of $\pi$.

The key point of this equivalence is that both the difference 
of group velocities of the mass eigenstates and the difference of the  
oscillation phases for different energies in the WP are determined by 
the same quantity $d(\Delta H_m) / dE$.  
According to \Eqref{eq:LCohMomSpaceVar} and (\ref{eq:sepgen}),
\begin{equation}
	\Delta x_\text{shift} = -\frac{\partial\phi}{\partial E} \;,
\label{eq:sep-der}
\end{equation}
which holds for arbitrary $L \neq L_\text{coh}$, too.

A zero value of $\Delta x_\text{shift}$, i.e., no shift of the WP,
corresponds to a zero derivative $\partial\phi/\partial E$.
This implies a weak dependence of the oscillation phase
on energy in a certain interval, consequently no significant effect
of averaging, and thus no decoherence.
However, higher-order terms in the Taylor expansion of the oscillation
phase lead to $\Delta\phi \neq 0$ even if $\partial\phi/\partial E$
vanishes.  Therefore, the consideration in momentum space implies that
the coherence length increases significantly for
$\partial\phi/\partial E = 0$ but does not become infinite, which is
different from what we observed using the configuration-space treatment
in \Secref{sec:PropDec}.

\subsection{Impact of the detection process}
%%%%%%%%%%%%%%%%%%%%%%%%%%%%%%%%%%%%%%%%%%%%%%%%%%%%%%%%%%%
\label{sec:LCohMomSpace1}

In the adiabatic case considered so far, loss of WP overlap has no effect in practice,
once the energy resolution of the detector $\Delta E$  is taken into account. Indeed,  in
order to observe the oscillation pattern, $\Delta E$ has
to be sufficiently small to satisfy
\begin{equation} \label{eq:EResolutionDetector}
	\left| \phi(E-\Delta E,L) - \phi(E+\Delta E,L) \right| < 2\pi \;,
\end{equation}
which is analogous to \Eqref{eq:LCohMomSpace} 
with $\sigma_E$ replaced by $\Delta E$ and in fact suffices to
guarantee the observation of oscillations even if \Eqref{eq:LCohMomSpace}
is violated.
In configuration space a small energy uncertainty $\Delta E$
implies a large time uncertainty $\Delta t$.  All WP arriving
within this time interval will be detected coherently and hence 
their effects in the detector can
interfere even if they do not overlap.  In this way, the detector
\emph{restores} coherence \cite{Kiers:1995zj}.

If inequality \eqref{eq:EResolutionDetector} is violated, averaging the oscillation
probability over the energy interval $[E-\Delta E, E+\Delta E]$ destroys
the oscillation pattern just like the separation of the WP or
averaging over the WP size in momentum space.  
As a short cut, we could take into account the detection
process by introducing a generalized WP with width $\sigma_{E,\text{tot}}$
such that \cite{Giunti:1997sk}
\begin{equation}
	\frac{1}{\sigma_{E,\text{tot}}^2} =
	\frac{1}{\sigma_E^2} + \frac{1}{\Delta E^2} \;.
\label{eq:genwidth}
\end{equation}

%%%%%%%%%%%%%%%%%%%%%%%%%%%%%%%%%%%%%%%%%%%%%%%%%%%%%%%%%%%%%%%%%%%
\subsection{Equivalence of wave packet separation and energy averaging}
%%%%%%%%%%%%%%%%%%%%%%%%%%%%%%%%%%%%%%%%%%%%%%%%%%%%%%%%%%%%%%%%%%%%%%%

Essentially, \Eqref{eq:LCohMomSpace}  means that $2\sigma_E$ equals
the period of the oscillatory pattern in energy when $L = L_\text{coh}$,
which we denote by $E_T (L_\text{coh})$.  
Then from \Eqref{eq:cohener} and (\ref{eq:sepgen})
we obtain $|\Delta x_\text{shift}| = \pi/\sigma_E$ or
\be
	E_T = \frac{2\pi}{|\Delta x_\text{shift}|} \;.
\label{eq:etxs}
\ee
As one can easily verify, \Eqref{eq:etxs} holds for any baseline $L$. 
It is this relation between the period of the oscillatory curve in energy and the shift 
of the WP in configuration space that ensures equivalence 
of results obtained 
in the momentum and configuration space considerations.
One could say equally well that the equivalence is
due to \Eqref{eq:sep-der}, without having to introduce $E_T$.  

In the case of complete overlap,
$\Delta x_\text{shift} \rightarrow 0$,
the relation (\ref{eq:etxs}) gives $E_T \rightarrow \infty$,
which is equivalent to ${\partial\phi}/{\partial E} \rightarrow 0$.
In this case the effect of the interference term
(deviation from the averaged probability) does not depend
on energy, and therefore it is also independent of the energy resolution
of the detector (as long as $\Delta E \gg E$, as required by the Taylor
expansion in eq.~\ref{eq:LCohMomSpaceVar}).

In the opposite case of large shift $\Delta x_\text{shift}$,
the period becomes very small, so one needs to have
very good energy resolution since the condition $\Delta E \ll E_T$
should be satisfied to observe the interference
(oscillatory effect) in the oscillation probability.
In configuration space that would correspond
to a long coherent observation time  with $\Delta t \sim 1/ \Delta E$, and consequently
to restoration of coherence in the detector.

Summarizing the two pictures, in both representations (configuration and
momentum space) we start from the eigenvalues of the Hamiltonian
$H_{im}$ and their difference $\Delta H_m$.
In configuration space $\Delta H_m$ determines the 
difference of group velocities
and the relative shift of the WP\@. Then comparing
the shift with the effective size of the packet
$\sigma_{E,\text{tot}}^{-1}$
(which includes both the produced WP and the energy resolution of the
detector) determines whether coherence is preserved or lost.
In momentum space $\Delta H_m$ determines
the oscillation phase and the oscillatory
period in energy $E_T$.  Comparison of the latter with
the effective width of the packet $\sigma_{E,\text{tot}}$ 
determines whether the oscillatory pattern is observable or averaged to
a constant oscillation probability.

Equivalence of the configuration and momentum space considerations is
realized when the whole process is taken into account:
production,  propagation and detection  of neutrinos. 
The phase $\phi(E, L)$ is the key (integral) characteristic
which takes into account all the relevant (for coherence)
features of propagation.

The discussion up to this point shows that WP separation and
energy averaging produce equivalent effects in the adiabatic case.
In fact, this also follows from theorems in
\cite{Kiers:1995zj,Stodolsky:1998tc}, according to which it is
impossible to distinguish long and short WP; in particular, whether one
can observe coherent effects or not is independent of the size of the WP\@.
In the following, we will consider neutrino oscillations in matter with
density jumps, aiming to show explicitly that the
equivalence holds under such conditions as well.

\section{Coherence in multi-layer medium}
\label{sec:Cohandaddv}
%%%%%%%%%%%%%%%%%%%%%%%%%%%%%%%%%%%%%%%%%%%%%%%%%%%%%%%%%

The picture described in the previous section is modified if adiabaticity is broken. 
In what follows we will consider special (maximal) adiabaticity breaking 
occurring when neutrinos propagate in a multi-layer medium
that consists of several layers with constant or
adiabatically changing density and abrupt density changes between the layers.
In other words, there is a step-like change or jump of density at each
border between two layers. 
This happens in a supernova at the front of the shock waves.
Later, neutrinos experience  density jumps when they enter the Earth and
at the boundary between the mantle and the core.
Coherence in such a case can be treated in the same way as before,
taking into account splittings of the eigenstates at the borders.

\subsection{Splitting of eigenstates}
\label{sec:splittt}
%%%%%%%%%%%%%%%%%%%%%%%%%%%%%%%%%%%%%%%%%%%%%%%%%%%%%%%%%%%%%%%%%%%%%%%%

Let us consider the jump of the density between the layers $k$ and 
$k+1$. Suppose a neutrino propagates in the layer $k$, 
crosses the border and then propagates in the layer $k + 1$.
The eigenstate $\nu_{i m}^{(k)}$ in the layer $k$ does not coincide 
with any eigenstate in the layer $k + 1$. Therefore, when crossing 
the border, $\nu_{i m}^{(k)}$
will split into two%
\footnote{In the case of two-neutrino mixing.}
eigenstates $\nu_{j m}^{(k+1)}$ of layer $k + 1$.
Correspondingly, at a density jump each WP splits up into a pair of new
packets.  After the split, the state will oscillate and the packets will
shift according to the group velocity difference in the second layer.

Crossing a medium with $n$ layers, $2^n$ components of the neutrino
state are generated.  They correspond to parts of the WP of
$\nu_{1m}^{(n)}$ and $\nu_{2m}^{(n)}$ with different shifts.  Thus,
compared to the adiabatic case where one deals with only two WP that
either overlap or do not, there are additional possibilities in the multi-layer 
medium.
It is possible that some but not all WP overlap in the detector,
allowing to observe \emph{a part} of the interference terms in the
oscillation probability.  This corresponds to an intermediate case
between complete coherence (all WP overlap) and complete decoherence (no
WP overlap). Notice that the splitting has sense only in the presence of 
shift and separation of the WP\@. If the shift is neglected in each layer we can 
sum up the components which belong to the same eigenstate and the picture 
is reduced again to the propagation of two WP.   

The splitting of eigenstates at a density jump corresponds to the
decomposition
\begin{equation} \label{eq:StateSplitting}
	\begin{pmatrix} \nu_{1m}^{(k)} \\ \nu_{2m}^{(k)} \end{pmatrix} =
	\begin{pmatrix} c_{k+1,k} & s_{k+1,k} \\ -s_{k+1,k} & c_{k+1,k} \end{pmatrix}
	\begin{pmatrix} \nu_{1m}^{(k+1)} \\ \nu_{2m}^{(k+1)} \end{pmatrix} ,
\end{equation}
where
\begin{equation}
c_{k+1,k} \equiv \cos (\theta_{k+1} - \theta_{k})
\quad,\quad
s_{k+1,k} \equiv \sin (\theta_{k+1} - \theta_{k})
\end{equation}
are the cosine and sine of the change of the mixing angle at the jump,
\begin{equation}
	\theta_{k+1,k} \equiv \theta_{k+1} - \theta_{k} \;,
\end{equation}
and $\theta_{k}$ is the mixing angle in matter in the layer $k$.
If the matter density varies adiabatically within the layers, we use
$\theta_{k+1,k} \equiv \theta_{k+1}^\text{(i)} - \theta_k^\text{(f)}$,
the difference of the mixing angle at the beginning of layer $k+1$ and
the one at the end of layer $k$.
The vacuum mixing angle is $\theta_0 \equiv \theta$.
We will also use $c_k \equiv \cos\theta_k$ and $s_k \equiv \sin\theta_k$.

For the difference of mixing angles we find
\be
\sin 2 (\theta_{k+1} - \theta_{k}) =
\frac{\xi_{k+1}-\xi_{k}}{R(\xi_{k+1}) R(\xi_{k})} \, \sin 2\theta \;,
\label{eq:diffangle}
\ee 
where $R(\xi)$ is defined in \Eqref{eq:resfactor}.
In the case of small densities, $|\xi_k| \ll 1$, we have 
\be 
\sin 2 (\theta_{k+1} - \theta_{k}) \simeq
(\xi_{k+1} - \xi_{k}) \, \sin 2\theta \;.
\label{eq:smalld}
\ee

\subsection{Two layers of matter and catch-up effect}
\label{sec:2Layers}
%%%%%%%%%%%%%%%%%%%%%%%%%%%%%%%%%%%%%%%%%%%%%%%%%%%%%%%%%%%%%%%%%%%

As a simple explicit example, let us consider vacuum followed by two
layers of matter with lengths $L_1$ and $L_2$, constant densities
$\rho_1$ and $\rho_2$ and the corresponding effective
mixing angles $\theta_1$ and $\theta_2$. At the end of the second layer
we place a detector sensitive to $\nu_e$.  
Suppose the mass eigenstate $\nu_1$ arrives at the border of the first layer.  
Here the state splits into a pair of new WP corresponding
to the eigenstates in matter with density~$\rho_1$,
\begin{equation}
	\nu_1 = c_{10} \nu_{1m}^{(1)} + s_{10} \nu_{2m}^{(1)} \;.
\end{equation}
The new eigenstates  $\nu_i^{(1)}$  propagate to the end of the layer, 
acquiring the oscillation phase $\phi_1 = \Delta H_m^{(1)} L_1$.
The abrupt density change from $\rho_1$ to $\rho_2$ transforms the
eigenstates in matter according to \Eqref{eq:StateSplitting} with $k=1$.
Before reaching the detector they acquire another oscillation phase
$\phi_2 = \Delta H_m^{(2)} L_2$.

Assuming complete
coherence, the probability for observing an electron neutrino
in the detector is then (for a single neutrino energy $E$)%
\footnote{The oscillation probability and the phases depend not only on
$E$ but also on $L_k$ and $\rho_k$, but we do not write these
dependences explicitly in the following.}
\begin{equation}
\label{eq:2layers}
	P_{\nu_1\to\nu_e}(E) =
	\left|
	c_2 c_{21} c_{10} - c_2 s_{21} s_{10} e^{i\phi_1} +
	s_2 s_{21} c_{10} e^{i\phi_2} + s_2 c_{21} s_{10} e^{i(\phi_1+\phi_2)}
	\right|^2 .
\end{equation}
Here we have projected the eigenstates in the second layer onto $\nu_e$ according to 
$\nu_e = c_2 \nu_{1m}^{(2)} + s_2 \nu_{2m}^{(2)}$.
The four terms in \Eqref{eq:2layers} correspond to the four 
components of the state after two splits. 
The interference terms are proportional to $\cos\phi_1$,  $\cos\phi_2$,
$\cos (\phi_1 + \phi_2)$,
and $\cos (\phi_1 - \phi_2)$, that is, to the cosines 
of all possible combinations of the two phases. 

\begin{figure}
\centering
\includegraphics[width=0.5\linewidth]{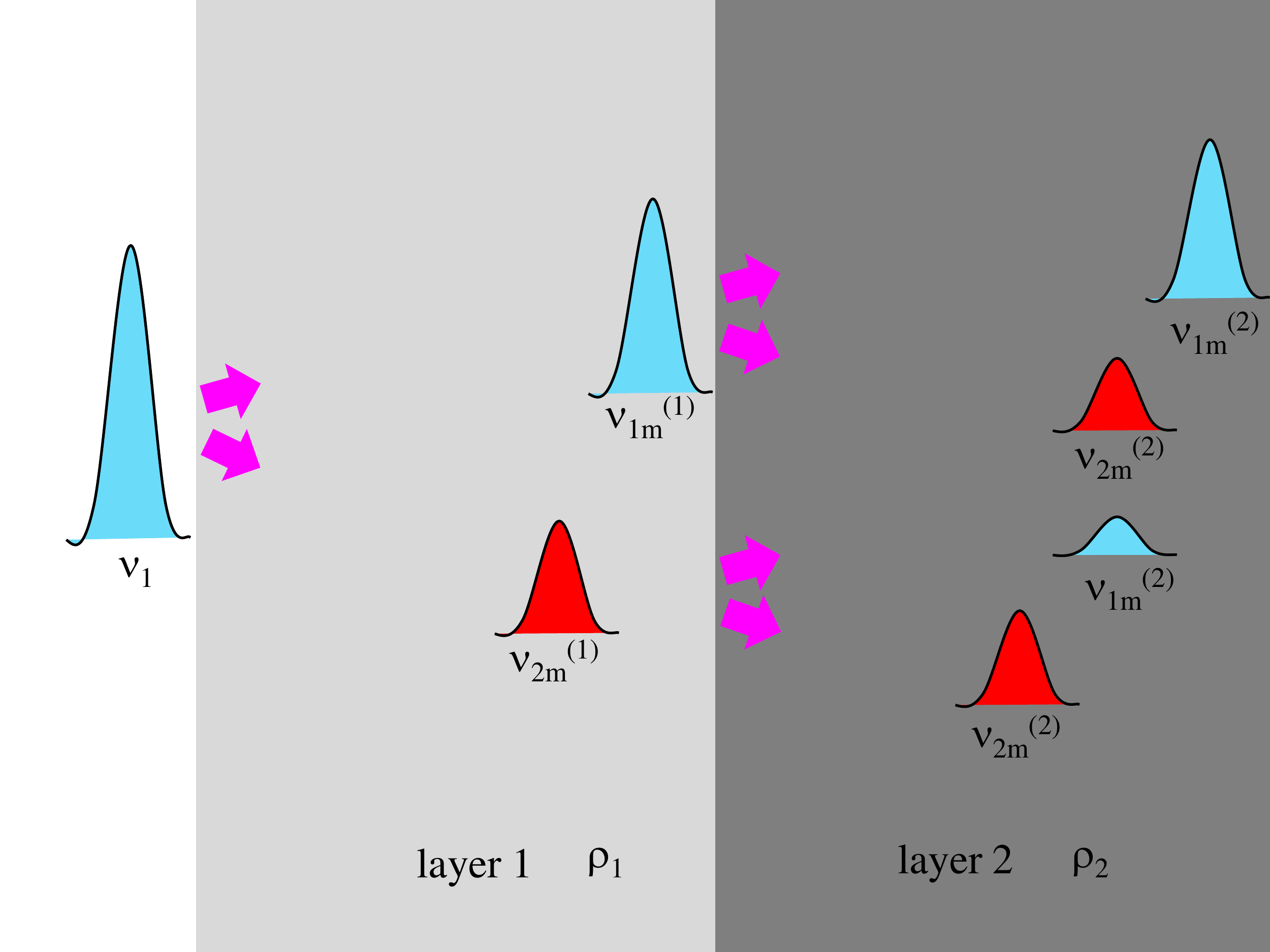}
\caption{Illustration of the splitting of the neutrino states and catch-up
effect in a medium with two layers. The wave packet
$\nu_{1m}^{(2)}$ catches up with the packet $\nu_{2m}^{(2)}$,
so that they start to overlap in the second layer.}
\label{fig:catchup}
\end{figure}

The splitting of WP at each boundary can lead to the particularly
interesting situation that although no WP overlap for some time during
the propagation, two WP overlap again when they reach the detector.
Suppose that $v_{1m}^{(k)} > v_{2m}^{(k)}$ in both layers.  Then
$\nu_{2m}^{(1)}$ falls behind during the propagation through the first
layer.  However, its splitting at the boundary creates a
$\nu_{1m}^{(2)}$ WP, which can \emph{catch up} with the $\nu_{2m}^{(2)}$
WP that originated from $\nu_{1m}^{(1)}$,
as illustrated in \Figref{fig:catchup}.
These two WP overlap in the detector, independently of their size, if
\begin{equation} \label{eq:CatchUp}
	\Delta v_m^{(1)} L_1 = \Delta v_m^{(2)} L_2 \;,
\end{equation}
where $\Delta v_m^{(k)}$ denotes the group velocity difference in layer
$k$.  Hence, the observed oscillation probability $P_{\nu_1\to\nu_e}$
will contain the corresponding interference term.  As $\Delta v_m^{(k)}$
depends on energy, the catch-up condition \Eqref{eq:CatchUp} can only be
satisfied for a particular energy, however.

This catch-up effect depends crucially on the WP picture in
configuration space.  At first sight, it seems unlikely to recover such
a complicated effect in the momentum space picture.  Thus, one might
hope that this effect breaks the equivalence between WP separation and
energy averaging, potentially allowing to construct a setup where the
observed oscillation probability depends on the WP size.  This is not
the case, though.
According to \Eqref{eq:sep-der},
\begin{equation} \label{eq:sep-derLayers}
	\Delta v_m^{(k)} L_k = \Delta x_\text{shift}^{(k)} = 
	-\frac{\partial\phi_k}{\partial E} \;.
\end{equation}
Therefore, \Eqref{eq:CatchUp} is equivalent to
\begin{equation} \label{eq:Stationarity1-2}
	\frac{\partial}{\partial E}(\phi_1-\phi_2) = 0 \;.
\end{equation}
This condition implies that one combination of oscillation phases,
$\phi_1-\phi_2$, depends only weakly on energy.  Then averaging 
$P_{\nu_1\to\nu_e}(E)$ over $\sigma_E$ or $\Delta E$ will not suppress
the interference term proportional to $\cos(\phi_1-\phi_2)$.
Consequently, the observed oscillation probability will contain this
term.  Once again, the considerations in configuration and momentum
space lead to the same result.

Let us study energy averaging more systematically in order to
find out how much the coherence length can grow due to the catch-up
effect and to interpret the remaining combinations of
oscillation phases as well.  The experimentally observed oscillation
probability is given by the energy average
\begin{equation} \label{eq:EnAverage}
	\overline{P(E)} =
	\int dE' \, P(E') \, f(E') \;,
\end{equation}
where $f(E)$ is the energy resolution function of the detector.
As a first approximation, we can replace $f(E)$ by a step function that
is non-zero only in the interval $[E-\Delta E, E+\Delta E]$, where
$\Delta E$ is the energy resolution.

Let us consider one combination of oscillation phases and denote it by
$\psi(E)$.  Analogously to \Eqref{eq:EResolutionDetector},
$\cos\psi$ in the corresponding interference term survives
energy averaging as long as
\begin{equation} 
\label{eq:LCohMomSpaceMultiLayers}
	|\Delta\psi(E)| =
	\left| \psi(E-\Delta E)-\psi(E+\Delta E) \right| < 2\pi \;.
\end{equation}
The Taylor expansion of $\psi$ yields
\begin{equation} \label{eq:TaylorPsi}
	\Delta\psi(E) =
	-2E \left.\frac{\partial\psi}{\partial E}\right|_E \frac{\Delta E}{E} -
	 \frac{E^3}{3} \left.\frac{\partial^3\psi}{\partial E^3}\right|_E
	  \left(\frac{\Delta E}{E}\right)^3 + \dots \;.
\end{equation}
Thus, the variation of $\psi$ around the energy satisfying
\begin{equation} \label{eq:MinDeltaPsi}
	\left.\frac{\partial\psi}{\partial E}\right|_E = 0
\end{equation}
is suppressed by two additional powers of $\Delta E/E$ compared to the
usual case.  Correspondingly, the propagation length can be larger before
$\Delta\psi$ reaches $2\pi$.%
\footnote{Alternatively, for a fixed baseline a larger $\Delta E$ is
sufficient to observe the interference term.}
In other words, the individual coherence
length for the $\cos\psi$ interference term increases considerably,
although it does not become infinite.  
In an experiment, this would show up as a deviation from the averaged
oscillation probability expected in the totally incoherent case for a
certain energy range and baseline.  One would not observe a complete
oscillation pattern in energy, since \Eqref{eq:MinDeltaPsi} can only
be satisfied for a particular energy.

Next, we consider the stationarity condition (\ref{eq:MinDeltaPsi}) for
different interference terms, i.e., different combinations $\psi$. 
\begin{enumerate}
\item For $\psi = \phi_k$ ($k = 1,2$), it can be satisfied only if
$\xi_k = \xi_0 = (\cos2\theta)^{-1}$ and thus $\Delta v_m^{(k)} = 0$,
i.e., the WP do not separate in layer $k$.  For this case we had found
the stationary-phase condition already at the end of
\Secref{sec:LCohMomSpace}.

\item For $\psi = \phi_1 - \phi_2$, we obtain \Eqref{eq:CatchUp}, as
already discussed.  Using (\ref{eq:Deltav}), we can write this equation as 
\begin{equation} \label{eq:cond12}
	L_2 =
	L_1 \, \frac{1-\xi_1\cos2\theta}{1-\xi_2 \cos2\theta} \,
	\frac{R(\xi_2)}{R(\xi_1)} \;.
\end{equation}
As $L_1$ and $L_2$ are positive, this condition can only be satisfied
if $\xi_1, \xi_2 < (\cos 2\theta)^{-1}$, or if
$\xi_1, \xi_2 > (\cos 2\theta)^{-1}$, that is, if both densities
are below or above the critical density.%
\footnote{Condition \eqref{eq:cond12} can also be satisfied for
arbitrary densities in the non-resonance channel, where $\xi_k<0$.} 
This means that either in both layers the eigenstates $\nu_{1m}$ move
faster than $\nu_{2m}$, or in both layers $\nu_{1m}$ move more slowly
than $\nu_{2m}$.  The overlap occurs in the second layer between the WP
of $\nu_{2m}^{(2)}$ and $\nu_{1m}^{(2)}$ originating from the transitions
\[
	\nu_1 \to \nu_{1m}^{(1)} \to \nu_{2m}^{(2)}
	\quad\text{and}\quad
	\nu_1 \to \nu_{2m}^{(1)} \to \nu_{1m}^{(2)} \;,
\]
where the arrows indicate the transitions at the density jumps.  The
corresponding interference term that is not averaged to zero arises from
the second and third terms in the oscillation probability
(\ref{eq:2layers}).  According to \Eqref{eq:2layers} the oscillation
depth for this mode is
$4 s_2 c_2 s_{10} c_{10} s_{21}^2 =
 \sin 2\theta_2 \, \sin 2\theta_{10} \, s_{21}^2$. 

\item For $\psi = \phi_1 + \phi_2$, we obtain from \Eqref{eq:MinDeltaPsi}
using \Eqref{eq:sep-derLayers} and (\ref{eq:Deltav}) 
\begin{equation} \label{eq:cond1+2}
	L_2 =
	-L_1 \, \frac{\Delta v_m^{(1)}}{\Delta v_m^{(2)}} =
	-L_1 \, \frac{1-\xi_1\cos2\theta}{1-\xi_2 \cos2\theta} \,
	\frac{R(\xi_2)}{R(\xi_1)} \;.
\end{equation}
This condition can be satisfied if $\xi_2 > (\cos 2\theta)^{-1} > \xi_1$,
or if $\xi_1 > (\cos 2\theta)^{-1} > \xi_2$, that is, if one density is
larger than the critical density and the other one is smaller.
Now either $\nu_{1m}^{(1)}$ moves faster than $\nu_{2m}^{(1)}$ whereas
$\nu_{1m}^{(2)}$ moves more slowly than $\nu_{2m}^{(2)}$, or vice versa.
The overlapping WP originate from the transitions
\[
	\nu_1 \to \nu_{1m}^{(1)} \to \nu_{1m}^{(2)}
	\quad\text{and}\quad
	\nu_1 \to \nu_{2m}^{(1)} \to \nu_{2m}^{(2)} \;.
\]
\end{enumerate}

In the limit of large densities, $|\xi_k| \gg 1$, the condition (\ref{eq:cond12})
reduces to
\be
L_2 \simeq L_1 \left[ 1 - \frac{\sin^2 2 \theta}{\cos 2\theta} 
\left(\frac{1}{\xi_1} - \frac{1}{\xi_2}\right) \right] .
\label{eq:cond12big}
\ee
For small densities, $|\xi_k| \ll 1$, we obtain 
\begin{equation} \label{eq:cond12small}
	L_2 \simeq L_1 \left[ 1 -
	 \frac{\sin^2 2\theta}{2} \left( \xi_1^2 - \xi_2^2 \right) \right] .
\end{equation}
So $L_2 \simeq L_1$ in both limits.  In addition, for small densities
the corrections are proportional to the square of the small parameters
$\xi_k$, i.e., strongly suppressed.

The case of two layers of matter with adiabatically varying density 
and density jump between them  can be realized in a supernova.
The jump is due to the shock wave.  
The new eigenstates propagate adiabatically and encounter the shock wave in
the MSW transition region, where the change of mixing angle is large 
(the effect of the shock wave outside this region is very small).
The first layer is between inner parts of the collective effects 
and the shock front, and the second one is between the shock front and
the surface of the star.
The oscillation phases $\phi_k$ should be computed for
adiabatically varying density,
$\phi_k = \int_k dx \, \Delta H_m^{(k)}$. 

The  possibility of two layers and two jumps is 
realized when there are two shock wave fronts (one can move inward),
see, e.g., \cite{Xu:2014tka}.  In this case the first layer 
is the one between the shock fronts and the second one is above the outer shock. 
The result is described by \Eqref{eq:2layers} with similar 
correspondence as in the first case.

\subsection{Generalization} \label{sec:PhaseCancellation}
%%%%%%%%%%%%%%%%%%%%%%%%%%%%%%%%%%%%%%%%%%%%%%%%%%%%%%%%%%%%%%%%%%%

The analysis can be immediately generalized to the case of $n$ layers
of matter.  In this case $n$ phases $\phi_k$ appear, and the amplitude
for flavor transitions can be written as 
\begin{equation}
	\mathcal{A}(E) = \sum_r a_r e^{i\psi_r} \;,
\label{eq:gena}
\end{equation}
where $\psi_r$ is any possible sum of the oscillation phases $\phi_k$ 
acquired in the individual layers,
$\psi_r = \{ 0, \phi_1, \phi_2, \dots, \phi_n,
\phi_1 + \phi_2, \dots, \phi_1 + \phi_n, \dots,
\phi_1 + \phi_2 + \phi_3, \dots \}$, and
$a_r$ are real numbers depending on energy only through
the energy dependence of the mixing angles in matter.
The number of phase combinations is $2^n$.  If we start from a single
mass eigenstate, the number of terms in \Eqref{eq:gena} is also $2^n$. 

The oscillation probability for a single energy is
\begin{equation}
	P(E) = |\mathcal{A}(E)|^2 \;.
\end{equation}
For two layers and observation of $\nu_e$ it is given explicitly in
\Eqref{eq:2layers}.  The probability contains interference terms
depending on the cosines of all possible combinations of the phases.
An experiment measures the energy-averaged probability $\overline{P(E)}$
defined in \Eqref{eq:EnAverage}.
We can distinguish three limiting cases.%
\footnote{To simplify the discussion we assume that there is a sharp
transition from coherence to decoherence, i.e., interference terms are
either present in $\overline{P(E)}$ or disappear completely.  Thus, we
neglect that interference terms are suppressed
but non-zero for a partial overlap of WP.}
\begin{enumerate}
\item If the detector's energy resolution is good enough
to resolve  all interference terms,
\begin{equation}
	\overline{P(E)} \simeq P_\text{coh} = |\mathcal{A}(E)|^2 \;.
\end{equation}
Coherence is completely preserved.
As discussed in \Secref{sec:LCohMomSpace1}, it does not matter whether
this happens because all WP overlap in the detector or because the
detector restores coherence.

\item For a bad resolution, the oscillation probability is given by the
incoherent sum
\begin{equation}
	\overline{P(E)} \simeq P_\text{decoh} = \sum_r a_r^2 \;. 
\end{equation}
Coherence is completely lost.
All WP are separated (and the detector does not restore coherence), or
averaging due to the bad energy resolution removes even interference
terms corresponding to overlapping WP.

\item It is possible that some but not all interference terms in the
probability survive averaging, for instance,
\begin{equation}
	\overline{P(E)} \simeq \sum_r a_r^2 + 2 a_s a_t \cos(\psi_s-\psi_t) \;.
\end{equation}
We will refer to this case as \emph{partial survival of coherence}.
Survival of coherence is possible for each interference term.
Therefore, we have to introduce an \emph{individual coherence
length} for each term.
\end{enumerate}

In the configuration-space picture, there are different reasons for
partial survival of coherence, depending on which type of interference
term survives.
On the one hand, if it contains a combination of several phases, the
survival is a consequence of the catch-up effect discussed in the
previous subsection.
In this case we can also speak of \emph{partial restoration} of
coherence, since the corresponding WP may cease to overlap for some time
before the catch-up effect restores the overlap.
The survival of terms with more than two phases corresponds to WP that
have different speeds in more than two layers before they meet again.
For more than two layers, it is possible that the catch-up effect causes
more than two WP to overlap in the detector.  In such a case, partial
restoration of coherence occurs for several interference terms at the same
time.

On the other hand, if the term containing $\cos\phi_k$ survives, this
can be due to a vanishing velocity difference in one layer,
$\Delta v_m^{(k)} = 0$, or due to the detector restoring the coherence
of consecutive WP that were separated in layer $k$.
Of course, there is also the trivial possibility that one or more layers
are so thin that WP do not separate inside them.
Another trivial example of partial survival of coherence is realized in
the case of three-neutrino mixing in a single layer when at large enough
distances the oscillation modes due to $\Delta m_{31}^2$ and
$\Delta m_{32}^2$ are averaged to zero, whereas the mode due to the
small splitting $\Delta m_{21}^2$ is not.

Note that the catch-up effect described here relies on two ingredients,
matter effects and strong (maximal) adiabaticity violation.  It is thus different from
the increase of the coherence length that is possible if $\Delta v_m$
changes sign during adiabatic propagation
\cite{Mikheyev:1989dy,deHolanda:2003nj}.  In that case, there is no
splitting of the WP into many components, and only two WP arrive at the
detector.

\section{Spread of the wave packets} \label{sec:Spread}
\subsection{Separation and spread}
%%%%%%%%%%%%%%%%%%%%%%%%%%%%%%%%%%%%%%%%%%%%%%%%%%%%%%%%%%%%%%%%

In the course of propagation between a supernova and the Earth in vacuum,
two effects occur: (i) separation of the WP
and (ii) spread (deliquescence) of each WP\@. 
Indeed, variations of the group velocity
$v = p/ E$, where $E = \sqrt{p^2 + m^2}$,
can be written as 
\begin{equation}
	\Delta v = \frac{\partial v}{\partial (m^2)} \Delta m^2 +
	 \frac{\partial v}{\partial p} \Delta p = 
	- \frac{p}{2 E^3} \Delta m^2 + \frac{m^2}{E^3} \Delta p \;.
\label{velocitydisp}
\end{equation}
Here the first  term  is the 
difference  of the group velocities of  the mass 
eigenstates as given in \Eqref{eq:DeltavmEarth}.%
\footnote{Up to a sign arising from our definition of $\Delta v_m$, see
\Eqref{eq:diffvel}.}
This term is responsible for the relative shift and eventual
separation of the WP of the mass eigenstates.
The second term in \Eqref{velocitydisp}
is  the velocity dispersion  due to 
different momenta in the WP\@.  It is responsible for a \emph{spread} 
of the WP of an individual mass eigenstate. 
For $\Delta p = \sigma_E$ we obtain 
the width of the packet after travelling a distance~$L$,
\be
	\sigma_\text{spread} = \frac{m^2}{E^3} \sigma_E L \;.
\label{s-spread}
\ee
The spread depends on the absolute 
value of the mass, and  for $\sigma_E = 1\MeV$ we obtain
\begin{equation}
	\sigma_\text{spread} \simeq 7\m \left(\frac{L}{10\kpc}\right)
	\left(\frac{15\MeV}{E}\right)^3
	\left(\frac{m^2}{7.5 \cdot 10^{-5}\eV^2}\right) .
\end{equation}
Thus, the WP of the two heavier states arriving at the Earth have a
macroscopic size.  For the lighest state, the size can be microscopic.

The separation of two mass eigenstates at the distance $L$ between a
supernova and the Earth equals according to \Eqref{eq:DeltavmEarth}
\begin{equation} \label{eq:ShiftSNEarth}
	\Delta x_\text{shift} \simeq
	51\m \left(\frac{L}{10\kpc}\right) \left(\frac{15\MeV}{E}\right)^2
	\left(\frac{\Delta m^2}{7.5 \cdot 10^{-5}\eV^2}\right) .
\end{equation}
The ratio of spread and separation does not depend on distance and is
given by
\begin{equation}
	\frac{\sigma_\text{spread}}{\Delta x_\text{shift}} = 
	2 \left(\frac{m^2}{\Delta m^2}\right) \left(\frac{\sigma_E}{E}\right) .
\end{equation}
It is determined by the relative difference of momenta and 
masses. For a hierarchical mass spectrum ($m^2 \simeq \Delta m^2$ for the
heavier mass eigenstate) and $\sigma_E/E < 1/2$,
the separation is larger than the spread.
A mild degeneracy with $m_2^2 \gtrsim 8 \, \Delta m^2_{21}$ or an
inverted mass hierarchy are sufficient to obtain
$\Delta x_\text{shift} < \sigma_\text{spread}$
for $E=15\MeV$ and $\sigma_E=1\MeV$.
In the case of an inverted hierarchy we have  $m_2^2/ \Delta m^2_{21} 
\simeq \Delta m^2_{31}/ \Delta m^2_{21} \simeq 33$, so
$\sigma_\text{spread}/\Delta x_\text{shift} \simeq 4$
and the WP of the eigenstates separated by $\Delta m^2_{21}$ will never
cease to overlap.
As we will show, in spite of this overlap the condition for coherence 
is not changed.

\subsection{Spread and energy redistribution} \label{sec:Redistribution}
%%%%%%%%%%%%%%%%%%%%%%%%%%%%%%%%%%%%%%%%%%%%%%%%%%%%%%%%%%%%%%%%

The spread is related to a
certain energy redistribution within the packet in 
configuration space: $E$ becomes a function of coordinate,
with the highest energies in the front of the WP and the lowest energies in the back. 
This can be seen by dividing the original momentum range into $n$ small
intervals with
$p_{j+1} - p_j = \delta p = \sigma_E/n$.  The whole WP
with average momentum $\bar p$,
\be
\Psi(x, t) = \sum_j \int_{p_j}^{p_{j+1}} \frac{dp}{(2\pi)^{1/2}} 
f(p-\bar{p}) \, e^{ipx-iEt} \;,
\ee
is then given by a sum over $n$ WP, which we will call
the \emph{small WP}\@.
The momentum-space width of a small WP equals $\sigma_{E}/n$ and
its spatial size is $n/\sigma_{E} \sim n \sigma_{x}$.
Introducing the average momentum and energy in each interval,
$\bar{p}_j$ and $\bar{E}_j$, the group velocity of the
small WP is $v_j = \bar{p}_j/\bar{E}_j$.
We choose $\delta p$ small enough to be able to neglect the spread of
the small WP against their original size for a given baseline
$L$.  The bigger the baseline, the smaller $\delta p$ has to be taken. 

The higher the $\bar{p}_j$ of a given small WP, the larger
the group velocity $v_j$.  The distance which such a WP propagates 
during time interval $t$ equals  
\be
x_j \simeq \left(1 - \frac{m^2}{2 \bar{p}^2_j}\right) t \;.
\label{xdd}
\ee
According to this, the small WP with higher 
$\bar{p}_j$ will be in front of those with smaller~$\bar{p}_j$.
Therefore, in the whole WP the front (forward edge) will have the
highest energy.  We denote the corresponding average momentum by $\bar{p}_n$.

The dependence of the average momentum $\bar{p}_j$  
on the distance from the front edge of the whole WP, $\Delta x_j$,
can be found in the following way. 
After propagation during the time~$t$, the position $x_n$ of the center
of the small (front) WP  with average momentum $\bar{p}_n$ 
is given by \Eqref{xdd} with $j=n$.  Consequently, the
distance (shift) between the packets $j$ and $n$ equals
\be 
\Delta x_j \equiv x_n - x_j \simeq \frac{m^2 t}{2}
\left(\frac{1}{\bar{p}^2_j} - \frac{1}{\bar{p}^2_n}\right)
\simeq \frac{m^2 L}{2}
\frac{\Delta p_j (\bar{p}_n + \bar{p}_j)}{\bar{p}_j^2 \bar{p}^2_n} \;,
\label{exxxx}
\ee
where $\Delta p_j \equiv \bar{p}_n - \bar{p}_j$ and
$L \simeq t$ is the baseline, the distance that the whole WP has
travelled.  Approximating $\bar p_n \simeq \bar p_j \simeq \bar p$ in
the sum and product of momenta,
which is justified for $\sigma_E \ll \bar p$,
we can rewrite \Eqref{exxxx} as 
\begin{equation} \label{eq:Deltaxint}
\Delta x_j
\simeq \Delta p_j L \, \frac{m^2}{\bar{p}^3} \;.
\end{equation}
In particular, for $j=1$ we obtain the spatial size of the whole WP,
leading to
$\sigma_\text{spread} \simeq \sigma_E L \, m^2/\bar p^3$,
which reproduces the result in \Eqref{s-spread}.
This confirms the validity of the presented picture. 
Thus, the spread of the whole WP can be described 
as separation of the small WP in configuration space.
Using \Eqref{s-spread} we can express \Eqref{eq:Deltaxint} as
\be
\sigma_\text{spread} \, \Delta p_j \simeq \Delta x_j \, \sigma_E \;.
\label{maineq}
\ee

We will show now that it is this energy redistribution which keeps 
the coherence condition unchanged.
The phase difference between the WP of two mass eigenstates 
(the oscillation phase)
can be written as \cite{Akhmedov:2009rb}
\begin{equation}
\phi = (x - v t) \, \Delta p - \frac{\Delta m^2}{2 E} t \;.
\label{totph} 
\end{equation}
In the case of packets without spread  
the difference of the average momenta equals $\Delta p \simeq \Delta m^2 /2E$
and $x - v t \lesssim 
\sigma_x$. Consequently, the first term in \Eqref{totph}
is usually small.
Here we consider the case with large spread of the WP instead, when 
$x - v t$ can be much larger than $\sigma_x$.  Suppose 
we have two WP with nearly the same spread.  This is realized, for
example, for a mass eigenstate that spreads significantly on the way
from a supernova to the Earth and splits into two eigenstates in
matter upon entering the Earth; these two eigenstates then shift with
respect to each other while propagating in the Earth.  The additional
spread during the propagation inside the Earth is negligible. 

\begin{figure}
\centering
\includegraphics[width=0.5\linewidth]{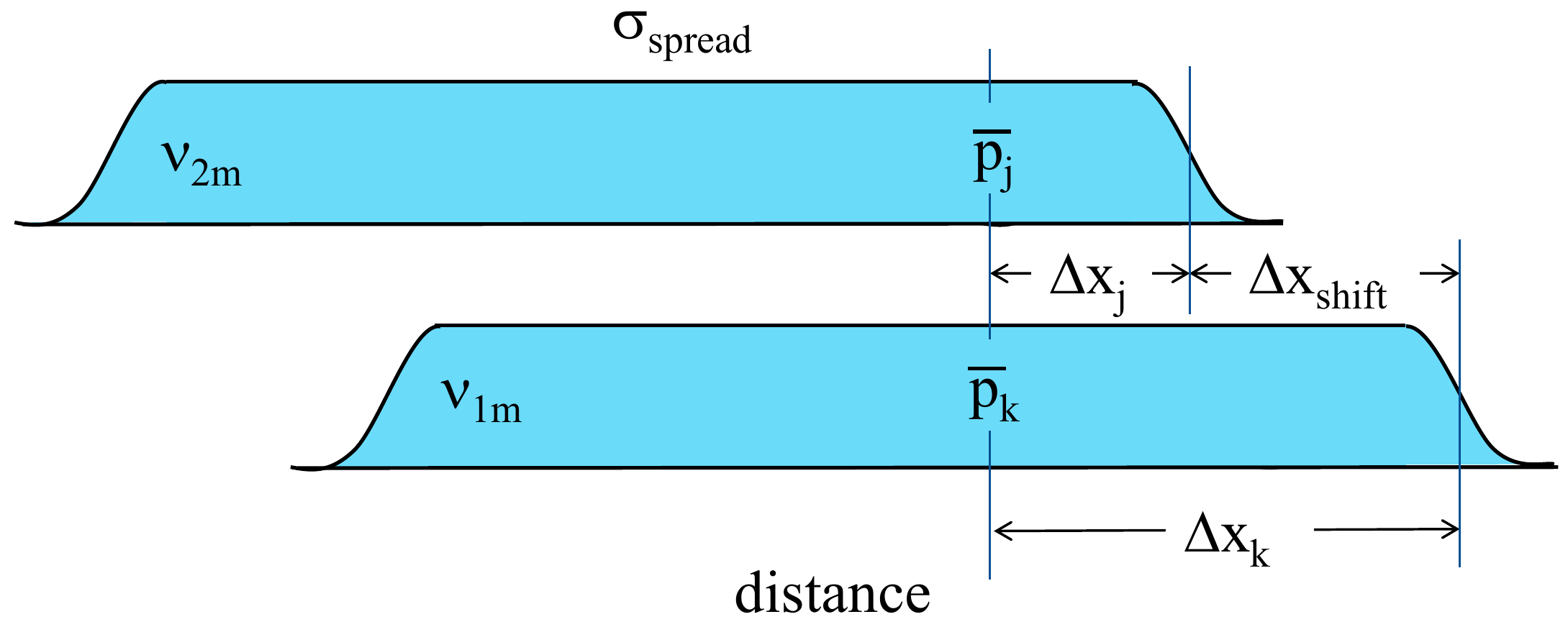}
\caption{Illustration of the shift of two wave packets with large spread.
Loss of coherence occurs even if the packets overlap due to the spatial
energy redistribution within the whole wave packets.
}
\label{fig:SpreadPackets}
\end{figure}

In order to calculate the oscillation probability we have to find the
phase difference of the two WP in the same space-time point.  We split
both WP into small WP as before.  Let us  consider the
space-time point where the small WP with average momenta
$\bar{p}_j$ and $\bar{p}_k$ are centered 
(the small WP with
$\bar{p}_j$ is a part of the slower WP, while the other small 
WP with $\bar{p}_k$ belongs to the faster WP), as illustrated in
\Figref{fig:SpreadPackets}.  Let us find the difference between
$\bar{p}_j$ and $\bar{p}_k$ arising from the
spatial shift of the WP\@.  According to \Eqref{maineq},
\be
\Delta p_j   = \Delta x_j \frac{\sigma_E}{\sigma_\text{spread}} \quad,\quad
\Delta p_k   = \Delta x_k \frac{\sigma_E}{\sigma_\text{spread}} \;.
\label{main-jk}
\ee
As $\Delta x_j$ and $\Delta x_k$ are the distances 
from the front parts of the packets,
\be
\Delta x_k = \Delta x_j + \Delta x_\text{shift} \;.
\label{diffkj}
\ee
Using \Eqref{main-jk} and \eqref{diffkj} we obtain 
\begin{equation}
\Delta p_{jk} \equiv \bar{p}_j - \bar{p}_k = \Delta p_k - \Delta p_j 
= \frac{\sigma_E}{\sigma_\text{spread}} \Delta x_\text{shift} \;.
\label{dlppp}
\end{equation}
This difference does not depend on $j$ and $k$.  That is, the difference
of momenta in the same space-time point is unchanged along the whole packets: 
$\Delta p = \Delta p_{jk}$.

Now $x - v t \simeq \pm\sigma_\text{spread}$ at the head and the tail of the
WP, and therefore neglecting the last term in 
\Eqref{totph} we obtain
\be 
\Delta\phi =
\phi(x_n)-\phi(x_1) \simeq
2\sigma_\text{spread} \Delta p = 2\sigma_E \Delta x_\text{shift}
\;.
\label{phirel}
\ee
Notice that $\sigma_\text{spread}$ cancels, 
which is related to the fact that a bigger spread implies a smaller gradient of $p$ 
and therefore a smaller $\Delta p$.
The condition for loss of coherence,
$|\Delta\phi| = 2\pi$, gives according to \Eqref{phirel}
\be
|\Delta x_\text{shift}| \, \sigma_E \simeq \pi \;.
\label{condspr}
\ee
Remarkably, this coherence-loss condition for packets with spread
coincides with the condition \eqref{eq:cohener} for short packets
without spread.  This coincides with the conclusion we
arrived at in energy-momentum space, thus confirming again the equivalence
of the two considerations. 
As a consequence, we can neglect WP spread when
discussing the coherence of supernova neutrinos oscillating in the
Earth.

The coherence-loss condition \eqref{condspr} does not depend on the absolute neutrino
mass scale, since $\sigma_\text{spread}$ cancels. It depends on $\Delta
m^2$ via $\Delta x_\text{shift}$.

\subsection{Spread of the wave packets in matter}
%%%%%%%%%%%%%%%%%%%%%%%%%%%%%%%%%%%%%%%%%%%%%%%%%%%%%%%%%%%

Let us consider the spread of WP in matter, generalizing the result 
(\ref{s-spread}). 
In general, according to \Eqref{eq:GroupVelocity}
\begin{equation}
\sigma_\text{spread} = \int_0^L dx \, \frac{d v_{im}}{d p} \Delta p =
\int_0^L dx \, \frac{d^2 H_{im}}{d p^2} \sigma_E \;,
\end{equation}
and the group velocities can be written as \cite{Mikheyev:1989dy}
\be
v_{im} = 1 - \frac{m_1^2 + m_2^2}{4 p^2} \mp \frac{1}{2} \Delta v_m \;,
\ee
where the upper sign ($-$) corresponds to $i=2$ and the lower one ($+$)
to $i=1$.
The difference of group velocities in matter, $\Delta v_m$, is given in
\Eqref{eq:Deltav}.  Differentiating with respect to $p$ and considering
a constant matter density for simplicity, we obtain 
\begin{equation}
\sigma_{i,\text{spread}} =
\frac{\sigma_E L}{2 p^3} \left[m_1^2 + m_2^2 \pm \Delta m^2
\frac{1 - 3\xi \cos 2\theta + \frac{3}{2} \xi^2 (1 + \cos^2 2\theta) - \xi^3 \cos 2\theta}%
{\left(1 - 2\xi\cos2\theta + \xi^2\right)^{3/2}} \right] .
\label{eq:spreadm}
\end{equation}

In the limit of small densities, $|\xi| \ll 1$, this expression reproduces 
the vacuum result~(\ref{s-spread}). In the opposite case of large densities,  
$|\xi| \gg 1$, \Eqref{eq:spreadm} gives
\begin{equation}
\sigma_{i,\text{spread}} \simeq \frac{\sigma_E L}{2 p^3}
\left( m_1^2 + m_2^2 \mp \sign(\xi) \, \Delta m^2 \cos2\theta \right) .
\end{equation}
In the case of large mixing, even if the spectrum is hierarchical the spread
of both packets will be comparable and given by the larger mass. 

The most interesting situation is the resonance region. 
For the critical value $\xi = 1/\cos 2\theta$, which corresponds 
to equal group velocities, $\Delta v_m = 0$, we obtain
\be
\sigma_{i,\text{spread}} =
\frac{\sigma_E L}{2 p^3} 
\left(m_1^2 +  m_2^2   \pm \frac{\Delta m^2}{2 \tan2\theta} \right). 
\label{eq:spr}
\ee
This can be rewritten in general  as 
\be
\sigma_{i,\text{spread}} =
\frac{\sigma_E L}{2p^3} \left[
(1 \mp a) \, m_1^2 + (1 \pm a) \, m_2^2 \right] .
\label{eq:gen-spr}
\ee
If $a > 1$,  then for one of the WP we can obtain
$\sigma_{i,\text{spread}} < 0$, which means that the WP \emph{shortens}
because its components with 
smaller energies move faster than the high-energy components.
In the case of a normal mass hierarchy,
$m_2^2 \gg m_1^2$, this happens for the WP of the lighter mass
eigenstate, for which
$\sigma_{1,\text{spread}} \simeq - (\sigma_E L/2p^3) (a - 1) m_2^2$.   
If the density varies on the way of the neutrinos, 
this may lead to the interesting phenomenon 
that one of the packets first spreads and then shortens again. 
In the specific case of \Eqref{eq:spr},
\be
a = \frac{1}{2 \tan2\theta} \;,
\ee 
and $a \simeq 1.6$ for the 1-3 mixing, whereas $a < 1$ for the 1-2 mixing.

In the MSW resonance $\xi = \cos 2\theta$, so
\begin{equation}
\sigma_{i,\text{spread}} =
\frac{\sigma_E L}{2 p^3}
\left(m_1^2 + m_2^2 \pm \Delta m^2 \frac{1 + \sin^2 2\theta}{2 \sin 2\theta} \right) .
\label{eq:spread-res}
\end{equation}
Now
\be
a = \frac{1 + \sin^2 2\theta}{2 \sin 2\theta} \simeq 1.9
\ee
for the 1-3 mixing.  For the 1-2 mixing, the parameter $a$ is only
slightly bigger than $1$.
 
In the realistic situation of a supernova this may happen in the MSW region. 
However, most of the spread occurs in vacuum on the way to the
Earth, and the spread inside
the supernova can be neglected.

\section{Oscillations of supernova neutrinos inside the Earth}
\label{sec:EarthOsci}
%%%%%%%%%%%%%%%%%%%%%%%%%%%%%%%%%%%%%%%%%%%%%%%%%%%%%%%%%%%%%%%%%%%%%%
\subsection{Neutrino states at the surface of the Earth}
%%%%%%%%%%%%%%%%%%%%%%%%%%%%%%%%%%%%%%%%%%%%%%%%%%%%%%%%%%%%%%%%%%%%%%

On the way from a supernova to the Earth the coherence of mass eigenstates is lost. 
In configuration space the WP are separated by a macroscopic distance
given by \Eqref{eq:ShiftSNEarth}.
Due to spread, the size of the WP, $\sigma_\text{spread} \sim$ few
meters, is also macroscopic, but this does not prevent decoherence, as
shown in \Secref{sec:Redistribution}.
Furthermore, the coherence cannot be restored by the detector.
Being separated by, say,  $30\m$ the WPs arrive at the detector 
within a time interval of $10^{-6}\,\text{sec}$.  So the whole detector
would have to be in a coherent state during $t > 10^{-6}\,\text{sec}$.
This can be clearly seen in the energy representation. 
To restore coherence the energy resolution of the detector should be
smaller than the period of oscillations in energy, $E_T$.  
The latter is determined by the condition
$|d\phi/dE| \, E_T = 2\pi$,
which is similar to the coherence condition (\ref{eq:LCohMomSpace}) considered before. 
Taking the expression for the phase in vacuum we obtain from this equality 
\be
E_T = E \frac{l_\nu}{L} = 2.5 \cdot 10^{-15} \, E \;,
\ee
where $l_\nu = 4\pi E / \Delta m^2$ is the oscillation length. 
A resolution $\Delta E /E < 10^{-15}$ cannot be obtained. 
Thus, each WP of a mass eigenstate will evolve inside the Earth
independently and the results
of their interactions in a detector will sum up incoherently. 

As we have discussed in \Secref{sec:Cohandaddv}, entering the Earth each 
mass eigenstate splits into eigenstates $\nu_{im}$
in the matter of the Earth and oscillates. We neglect the presence of the third neutrino 
$\nu_{3m} \simeq \nu_3$ here. This state decouples from the rest of the system 
producing just a small (given by $\sin^2 \theta_{13}$) average oscillation result. 
So we will consider two-neutrino oscillations driven by the mass splitting 
$\Delta m^2_{21}$. 

The Earth matter density is relatively small, 
yielding for the difference of group velocities 
\begin{equation} 
\label{eq:DeltavmEarth1}
	|\Delta v_m| \simeq \frac{\Delta m^2_{21}}{2E^2} K \;,
\end{equation}
where $K \simeq 1$ for $\xi \ll 1$, which is realized at low energies. 
Using \Eqref{eq:Deltav} we obtain $K = 0.995$ and $0.985$ 
for $E = 30\MeV$ and $50 \MeV$, respectively.
For the core and  $E = 50 \MeV$ we find $\xi = 0.35$  
and $K = 0.932$. So,  even for relatively large 
$\xi$ the parameter $K$ is close to unity.

\subsection{The coherence condition in the Earth}  
%%%%%%%%%%%%%%%%%%%%%%%%%%%%%%%%%%%%%%%%%%%%%%%%%%%%%%%%%%%%%%%%%%%

The Earth is a low-density medium for supernova neutrinos with $E < 30\MeV$.
Then, using the vacuum value for the difference of group velocities 
(as a first approximation) 
we find for the  shift (separation) of the eigenstates in the Earth 
\be
\Delta x_\text{shift} \simeq L_E \, \Delta v =  L_E \frac{\Delta m^2}{2 E^2} 
= 3.3 \cdot 10^{-10}\cm \left(\frac{L_E}{10^4\km} \right) 
\left(\frac{15\MeV}{E}\right)^2 ,
\label{wospread}
\ee
which is much smaller than the size of the WP after spread:
$\Delta x_\text{shift} \ll \sigma_\text{spread}$.  
One may therefore wonder whether the spread of the WP on their
way to the Earth can prevent decoherence.  
This is not the
case, as one can most easily see in momentum space.  Here the WP
width $\sigma_E$ is not changed in the course of propagation.   
Consequently, the above derivation of the coherence length 
in momentum space always yields the same result, regardless of WP spreading.  
As we have shown in \Secref{sec:Redistribution},
a detailed analysis in configuration space 
leads to the same result as in momentum space.  Therefore, we can use the same expression 
for the coherence length as for the WP of width $\sigma_x$ without spread.
The equivalence of WP separation and energy averaging relies on an
integration over the
detection time, which is necessary because it is not known precisely
at which moment of time a neutrino is produced.
This is equivalent to the assumption of a stationary source needed for
deriving the theorems of \cite{Kiers:1995zj,Stodolsky:1998tc}
about the indistinguishability of long and short WP\@.
The equivalence does not hold in the hypothetical case that the times of
both production and detection of a single neutrino are measured.

Thus, in the low-density limit, $\xi \ll 1$, we can use \Eqref{eq:DeltaxSN} and
\eqref{eq:lcohvac} to estimate
\begin{equation}
	L_\text{coh} \simeq (800 - 1800) \km \left(\frac{E}{15\MeV} \right)^2 .
\label{eq:cohle}
\end{equation}
Consequently, $L_\text{coh}$ is of the same order of magnitude as the
Earth's radius (or the sizes of its core and mantle).

Using expressions for the length of  trajectories in different layers we can find regions 
of complete decoherence, partial decoherence and complete coherence in the 
$E$--$\cos\eta$ plane, where $\eta$ is the nadir angle.

For the mantle-crossing trajectories, $|\cos\eta| < 0.83$, the length is given by 
\be
L_M = D_E \, |\cos\eta| \;,
\label{eq:lm}
\ee 
where  $D_E = 12742\km$ is the diameter of the Earth. 
Then the condition $L_M = L_\text{coh}$ and expressions (\ref{eq:lm}) 
and (\ref{eq:cohle})  give the upper bound on energy of the coherence loss region: 
\be
E_\text{dec} = 
48.9 \MeV \left(\frac{\sigma_x}{2 \cdot 10^{-11}\cm}\right)^{1/2} |\cos
\eta|^{1/2} \;.
\ee

For the lower energy bound of the coherence region defined by the condition
$L = 0.1 \, L_\text{coh}$
we have  
\be
E = 150\MeV \, |\cos \eta|^{1/2} \;.
\ee

For the core-crossing trajectories the length of each layer of mantle equals 
\be
L_M = R_E |\cos \eta| -  \frac{L_C}{2} \;,
\label{eq:lmantle}
\ee
where $L_C$ is the length of trajectory in the core:
\be
L_C = 2 \sqrt{R_C^2 - R_E^2 \sin^2 \eta} \;.
\label{eq:lcore}
\ee
Here $R_C = 3570\km$ is the radius of the core. 
Now the  upper energy bound on  the decoherence 
region in the core: 
\be
E = 48.9\MeV \left(\frac{\sigma_x}{2 \cdot 10^{-11}\cm}\right)^{1/2}
\left[ \left(\frac{R_C^2}{R_E^2}\right) - \sin^2 \eta
\right]^{1/4} .
\ee
For the mantle layers 
\be
E = 34.7\MeV
\left[\cos \eta - \left(\frac{R_C^2}{R_E^2} - 
\sin^2 \eta \right)^{1/2} \right]^{1/2}. 
\ee
These estimates of energy borders of regions of coherence and decoherence
are valid with about 30\% accuracy due to uncertainties 
in our estimate of $\sigma_x$. 

\begin{figure}
\centering
\includegraphics[width=0.5\linewidth]{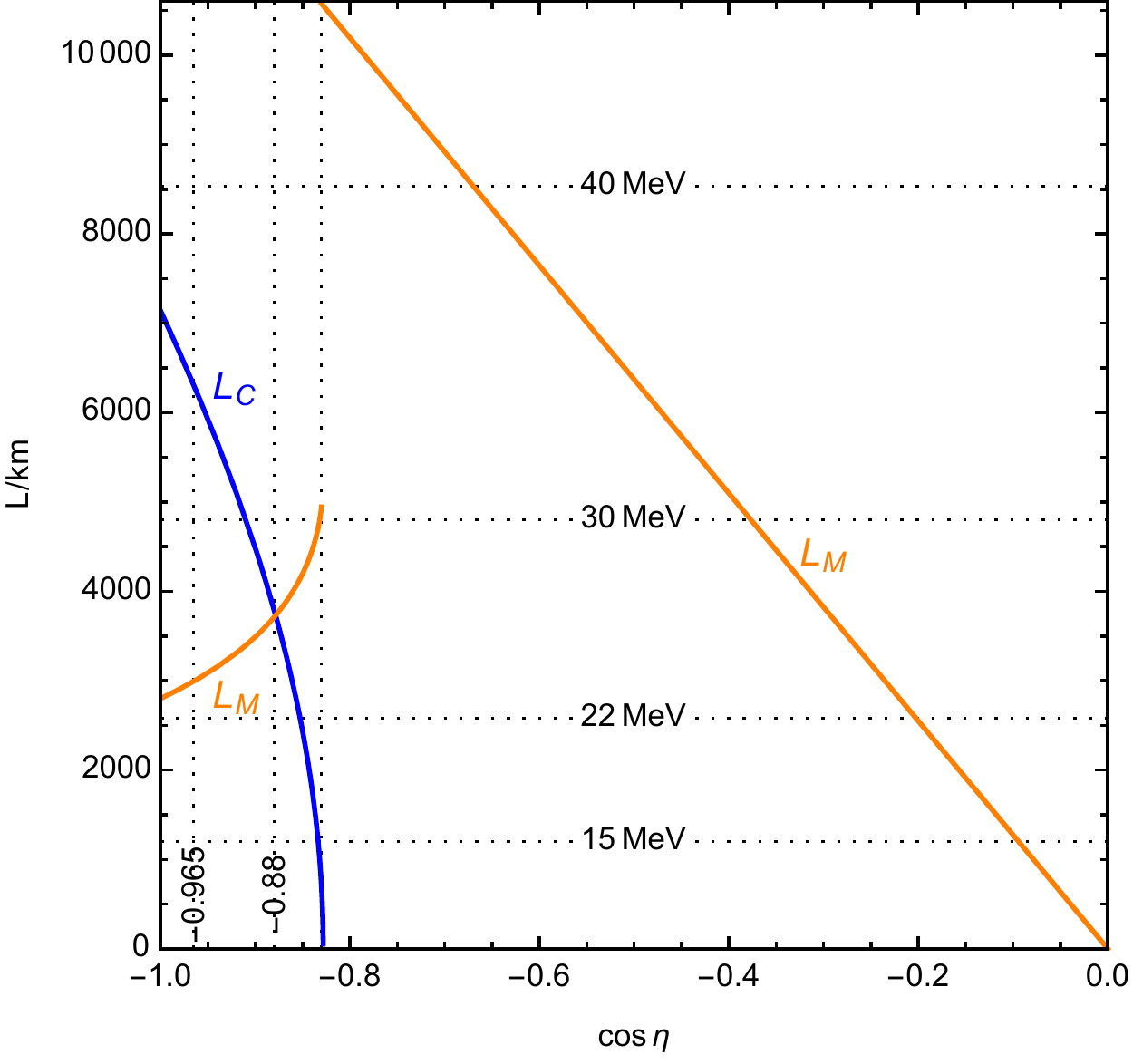}
\caption{Lengths of neutrino trajectories in different
layers of the Earth as functions of the nadir angle.
For $\cos\eta > -0.83$, $L_M$ is the length of the trajectory in the mantle.
For $\cos\eta < -0.83$, which corresponds to neutrinos crossing the core
and two mantle layers, $L_M$ is the length of the trajectory in a single
mantle layer.
Also shown are the coherence lengths for different
neutrino energies (horizontal dotted lines).
}
\label{fig:Trajectories}
\end{figure}

In \Figref{fig:Trajectories} we show the lengths of different trajectories 
in the mantle and the core as functions of nadir angle 
according to \Eqref{eq:lm}, (\ref{eq:lmantle}), and (\ref{eq:lcore}) together with  
the coherence lengths for different neutrino energies
(the horizontal lines). 

Notice that $\sigma_E \sim 1\MeV$ we have found does not depend 
on neutrino energy, whereas the energy resolution 
of the detector $\Delta E_\text{res} \propto \sqrt{E}$. 
We find that already at $E > (2-3) \MeV$ the energy resolution 
of a detector becomes larger than $\sigma_E$, and 
therefore determines the oscillation pattern apart from the 
cases of catch-up.

In what follows we will describe separately the effects of propagation 
along the mantle only and along the core-crossing trajectories.

\subsection{Oscillations in the mantle of the Earth}
%%%%%%%%%%%%%%%%%%%%%%%%%%%%%%%%%%%%%%%%%%%%%%%%%%%%%%%%%%%%%%%%%%%%%%%%%

We consider mixing of two neutrinos with mass splitting $\Delta m_{21}^2$. 
As we discussed before, incoherent fluxes of the mass eigenstates 
arrive at the Earth and we denote by $c_0^2$ and $s_0^2$ the fractions
of $\nu_1$ and $\nu_2$.
Neutrinos crossing only the Earth's mantle experience  one jump in
density upon entering the Earth. 
At this jump the mass states split into eigenstates in the Earth matter 
and start to oscillate.  
As we have established in \Secref{sec:splittt},  the effect of splitting
is determined by change the of the mixing angle due to the density jump. For
small densities it is given by \Eqref{eq:smalld},  or
\be
s_{ij} \simeq \frac{1}{2} (\xi_i - \xi_j) \, \sin 2\theta \;.
\label{eq:sijdif}
\ee
Here $\sin 2\theta = \sin 2\theta_{12} \simeq 0.925$.
Since propagation within layers is adiabatic 
one needs to use the values of $\xi$ immediately before and after a jump.
For the 0--1 jump $\xi_0 \simeq 0$, and $\xi_1$ is given by the density
at the surface of the Earth. Numerically,
\be
\xi_1 \simeq 0.0525 \left(\frac{E}{15\MeV} \right),  
\ee
and consequently, 
\be
s_{10} \simeq \frac{1}{2}\xi_1 \sin 2 \theta =   0.0243 \left(\frac{E}{15\MeV} \right).
\label{eq:s10}
\ee
The sine and split effect increase with energy. For $E = 30\MeV$ and 
$E = 60\MeV$ we have $s_{10} = 0.05$ and $0.1$, respectively. 

The density jump transforms the mass eigenstates into the eigenstates 
in the mantle according to
\begin{equation}
	\begin{pmatrix} \nu_1 \\ \nu_2  \end{pmatrix} =
	\begin{pmatrix} c_{10} & s_{10} \\ -s_{10} & c_{10} \end{pmatrix}
	\begin{pmatrix} \nu_{1m}^{(1)} \\ \nu_{2m}^{(1)} \end{pmatrix}.  
\end{equation}
Then the probability to find a $\nu_e$ is
\begin{equation}
P_\text{coh}(\nu_e) = 
c_0^2 \left| c_1 c_{10} + s_1 s_{10} e^{i\phi_1}\right|^2 +
s_0^2\left|s_1 c_{10} e^{i\phi_1} - c_1 s_{10}\right|^2 , 
\end{equation}
where 
\be
\phi_1 \simeq \frac{\Delta m_{21}^2}{2E} \int R(\xi(x)) \, dx
\ee
is the adiabatic phase acquired along the trajectory. 

If the WP of $\nu_{im}^{(1)}$ become separated,
we obtain 
\begin{equation}
	P_\text{decoh}(\nu_e) = 
	\left( c_1 c_0 c_{10} \right)^2 +
	\left( c_1 s_0 s_{10} \right)^2 +
	\left( s_1 c_0 s_{10} \right)^2 +
	\left( s_1 s_0 c_{10} \right)^2 .
\end{equation}
The difference between the complete coherence and loss of coherence  cases is
\begin{equation}
	P_\text{coh}(\nu_e) - P_\text{decoh}(\nu_e) = 
	\frac{1}{2} \cos 2\theta_0 \sin2(\theta_1-\theta_0) \sin 2\theta_1
	\cos\phi_1 \;.
\end{equation}
It vanishes when the phase $\phi_1$ is averaged, which occurs 
if  a detector has insufficient energy resolution to
observe the oscillation pattern. 

Inserting \Eqref{eq:s10} we obtain 
\be
P_\text{coh}(\nu_e) - P_\text{decoh}(\nu_e) 
\simeq  0.0243 \left(\frac{E}{15\MeV}\right)
\cos 2 \theta_0  \sin 2\theta \cos \phi_1 \;,
\ee
where we used that $\theta_1 \simeq \theta$.  
Then the depth of oscillations at $30\MeV$ can be 
about $5\% \cos 2 \theta_0$. In the case of normal mass ordering
\cite{Dighe:1999bi}
\be
\cos 2 \theta_0 = \frac{F_\mu - F_\tau}{F_\mu} \;.
\ee
If the fluxes of $\nu_\mu$ and $\nu_\tau$
which arrive at the region of MSW transitions are equal,    
$\cos 2 \theta_0 = 0$ and no oscillations are expected. 

For antineutrinos we have  
\be
\cos 2 \theta_0 = \frac{\bar{F}_e - 
\bar{F}_\mu}{\bar{F}_e} \;,
\label{eq:cos0}
\ee
and it is  expected to be small: of the order $0.1$. The fraction can be larger 
in the high-energy tail where the fluxes decrease
exponentially.  
In the case of inverted mass hierarchy the effect 
is zero in the antineutrino channel but non-zero in the neutrino 
channel with $\cos 2 \theta_0$ as in \Eqref{eq:cos0}
but with larger difference of the neutrino fluxes.  
So, one can expect  few $1 - 3 \%$ depth of oscillations.

%%%%%%%%%%%%%%%%%%%%%%%%%%%%%%%%%%%%%%%%%%%%%%%%%%%%%%%%%%%%%%%%%
\subsection{Oscillations of neutrinos crossing the core}
%%%%%%%%%%%%%%%%%%%%%%%%%%%%%%%%%%%%%%%%%%%%%%%%%%%%%%%%%%%%%%%%%

Neutrinos that cross the core of the Earth encounter three layers: 
mantle, core, and mantle.  Thus, they pass 3 jumps of density: 
at the surface when entering the Earth, 
when entering the core and when leaving the core. Therefore, each 
mass state arriving the Earth 
splits into $2^3 = 8$ components reaching  a detector. 

The change of mixing in the 0-1 jump is given in  
\Eqref{eq:s10}. For the 1-2 jump one should use the densities
at the surface of the core and in the deepest point of the mantle.
This gives
\be
s_{21} \simeq \frac{1}{2}\sin 2\theta (\xi_2 - \xi_1) \simeq 
0.0347 \left(\frac{E}{15\MeV} \right).
\ee
For $E = 30\MeV$ and $E = 60\MeV$ we obtain
$s_{21} = 0.07$ and $0.14$.  In the resonance region the
change of the angles is small since already the vacuum 1-2 mixing is large.
For larger energies
the size of the layers ($\sim 5000 - 6000\km$)
becomes smaller than the coherence length (\ref{eq:cohle}) and loss of coherence 
can be neglected (see \Figref{fig:Trajectories}).

As the two mantle layers are approximately symmetric,
for the mixing changes we have 
$c_{32} = c_{21}$, $s_{32} =  - s_{12}$,  
$\theta_1=\theta_3$ and for the phases $\phi_1 = \phi_3$. 
Then the probability of the $\nu_1 \rightarrow  \nu_e$ transition 
in three layers in the coherent case equals 
\begin{align}
	\!\!\!\!\!\!\!\!\!
	P_\text{coh}(\nu_1 \rightarrow \nu_e) &= 
	\left|
	 \left( c_{10} c_{21}^2 - 
          s_{10} s_{21} c_{21} e^{i\phi_1} + 
	  c_{10} s_{21}^2 e^{i\phi_2} 
        + s_{10} s_{21} c_{21}  e^{i (\phi_1 +  \phi_2)}
	\right) c_3 + {} \right.
\nonumber\\
& \hphantom{= \left| \right.} \left.  
	 \left( - c_{10} s_{21} c_{21} e^{i\phi_1} 
                + s_{10} s_{21}^2 e^{i 2 \phi_1} 
                 + c_{10} s_{21} c_{21} e^{i (\phi_1 +  \phi_2)}  
                 + s_{10} c_{21}^2 e^{i (2 \phi_1 + \phi_2)}
	 \right) s_3
	 \right|^2 ,
\label{eq:PcohEarth3}
\end{align}
where $c_3 = c_1$ is the cosine of the mixing angle in matter in layer~3.

Let us give an interpretation of the 8 terms
in \Eqref{eq:PcohEarth3}, which  correspond to 8 different channels 
(chains of transitions) and to  8 WP arriving at the detector.
The terms in the first line describe the whole  three layer transition 
$\nu_1 \to \nu_{1m}^{(3)} ( = \nu_{1m}^{(1)})$ with intermediate 
states in layers 1 (mantle) and 2 (core):  
\be 
\nu_{1m}^{(1)} \rightarrow \nu_{1m}^{(2)} \quad,\quad
\nu_{2m}^{(1)} \rightarrow \nu_{1m}^{(2)} \quad,\quad
\nu_{1m}^{(1)} \rightarrow \nu_{2m}^{(2)} \quad,\quad
\nu_{2m}^{(1)} \rightarrow \nu_{2m}^{(2)} \;.
\label{eq:inttr} 
\ee
The state $\nu_{1m}^{(3)} =  \nu_{1m}^{(1)}$ propagates in the third (mantle) 
layer being  then  projected onto $\nu_e$. 
The four terms in the second line of \Eqref{eq:PcohEarth3}
correspond to the whole transition $\nu_1 \rightarrow \nu_{2m}^{(3)} = \nu_{2m}^{(1)}$
with the same intermediate states and transitions in layers 
1 and 2  as in (\ref{eq:inttr}). 
Now the state $\nu_{2m}^{(3)} = \nu_{2m}^{(1)}$ propagates in the third 
layer being then projected onto $\nu_e$ in a detector.
Any transition $\nu_{2m}^{(i)} \leftrightarrow \nu_{1m}^{(j)}$ at the density jumps leads to 
the sine of the difference of the angles, $|s_{ji}|$, which is small.  
A minus sign is associated to the transitions 
$\nu_{2m}^{(1)} \rightarrow \nu_{1m}^{(2)}$ and  $\nu_{1m}^{(2)} \rightarrow \nu_{2m}^{(1)}$. 
Every appearance (propagation) of $\nu_{2m}^{(i)}$ in the chain 
of transitions leads to the oscillation phase $\phi_i$
since for definiteness we attach the phase factor to the state $\nu_{2m}^{(i)}$ 
in each layer. The state $\nu_{2m}^{(i)}$ (with larger mass) moves more
slowly than $\nu_{1m}^{(i)}$, 
it arrives later and has the bigger phase.  

The first term in \Eqref{eq:PcohEarth3} with 
the amplitude $c_{10} c_{21}^2$ and without phase factor corresponds to 
the fastest component of the state: in all three layers it corresponds to 
$\nu_{1m}^{(i)}$. Notice that this term has the largest amplitude. 
In contrast, the last term,  with the largest phase,  
$2 \phi_1 + \phi_2$  and the amplitude $s_{10} c_{21}^2$ corresponds 
to the slowest component when the heaviest component $\nu_{2m}^{(i)}$ 
propagates in all three layers.  
The WPs which correspond to these components 
cannot produce catch-up effect with other components. 

The interference terms  in the probability (\ref{eq:PcohEarth3})
are proportional to cosines of  all possible differences 
of the phases.  This includes $0$,  $\phi_1$, $\phi_2$,  
$\phi_1 + \phi_2$,  $2 \phi_1$,  $2 \phi_1 + \phi_2$, and 
 $\phi_1 - \phi_2$, $2 \phi_1 - \phi_2$ and the same combinations with 
opposite signs. 

Let us consider the interfence terms which can show up the ``catch-up'' effect. 
The critical energy (corresponding to $\xi_0$) is roughly $500\MeV$ in
the mantle and $200\MeV$ in the core.  So, the whole spectrum of
supernova neutrinos is below the critical energy, and consequently,   
$\xi_1, \xi_2 < \xi_0$. Furthermore, the average values of $\xi_1$ and $\xi_2$ 
are much smaller than $\xi_0$. In this case 
the derivatives of phases $d\phi_1/dE$ and $d\phi_2/dE$ are of the same signs 
as the signs the phases. Therefore, the cancellation of derivatives 
(which leads to the ``catch-up'' effect)   
can be realized for the interference terms with differences of phases:  
$0$, $\phi_1 - \phi_2$, $2 \phi_1 - \phi_2$. Explicitly, we obtain 
\begin{align}
	P_\text{coh}^\text{int}(\nu_e) & = 
	\frac{1}{4} \sin2\theta_{10} \sin^22\theta_{21} \sin2\theta_1 -
	s_{21}^3 c_{21} \sin2(\theta_{10}+\theta_1) \cos(\phi_1-\phi_2) + {}
\nonumber\\
&{} +
	\frac{1}{2} \sin2\theta_{10} s_{21}^4 \sin2\theta_1 \cos(2\phi_1-\phi_2) \;.
\label{eq:intcatch}
\end{align}
Note that in general the first term has the phase factor 
$\cos(\phi_1-\phi_3)$,  but in the symmetric case with $\phi_1 = \phi_3$,  
it equals 1.

\begin{figure}
\centering
\includegraphics[width=0.7\linewidth]{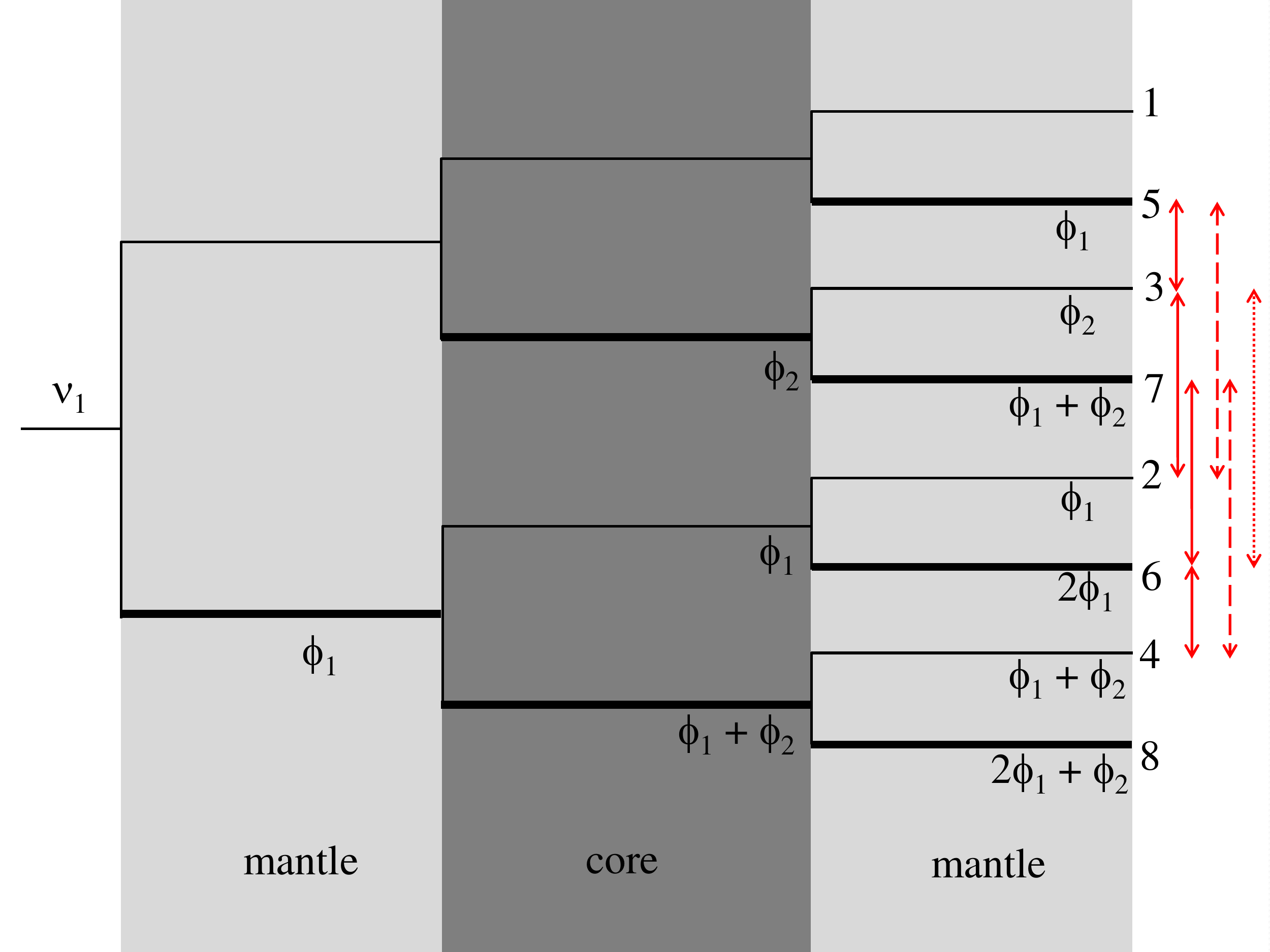}
\caption{Scheme of the neutrino state splitting in a medium
with three layers, which is realized for
core-crossing trajectories in the Earth.
Thin (thick) lines correspond to
components of states which propagate with
high (low) group velocity. Indicated are the phase
differences acquired by different components.
The numbers on the right-hand side indicate terms
in \Eqref{eq:PcohEarth3}.  The catch-up effect can occur
between states with equal phases. These states
are connected by vertical lines with arrows.
Solid, dashed, and dotted lines correspond to
the second, first, and third terms in \Eqref{eq:intcatch}.}
\end{figure}

Let us consider the three terms in \Eqref{eq:intcatch}
in detail, giving their interpretation in
configuration space. 

1) There are 4 contributions to the term 
with phase $(\phi_1-\phi_2)$.  All of them correspond to the catch-up 
in the case of 2 layers considered in \Secref{sec:2Layers}.  

(i) Interference of the 3rd and 5th terms of 
\Eqref{eq:PcohEarth3} corresponds to different motion of the WP in  
the second (core) and third (mantle) layers: the component $\nu_{1m}^{(1)}$ splits 
into $\nu_{1m}^{(2)}$ and $\nu_{2m}^{(2)}$  crossing the 
density jump between the mantle and the core. 
In the core $\nu_{1m}^{(2)}$ propagates faster than $\nu_{2m}^{(2)}$. 
Both  components split further  when entering the mantle again. 
Then $\nu_{1m}^{(3)}$, which originates from $\nu_{1m}^{(2)}$,
moves faster than $\nu_{2m}^{(3)}$, and therefore its WP can catch up with the packet of 
$\nu_{2m}^{(3)}$ originated from $\nu_{2m}^{(2)}$. Catch-up occurs in the third layer. 

(ii) Interference of 6th and 7th terms in \Eqref{eq:PcohEarth3} is similar 
to (i) with the only difference that 
the two interferring channels originate from  $\nu_{2m}^{(1)}$ (in the first layer). 

(iii) In the case of interference of 3rd and 2nd terms in \Eqref{eq:PcohEarth3} 
(in contast to (ii))  the first (mantle) and  the second (core) layers are involved. 
The catch-up occurs in the core;  in the third layer $\nu_{1m}^{(3)}$ 
propagates in both channels,  
so neither shift nor  phase difference are acquired. 
The consideration is similar  to the previous case.

(iv) Interference of 4th and 6th terms  is similar to 
case (iii) with the only difference that 
in the third layer in both channels $\nu_{2m}^{(3)}$ propagates. 

According to \Eqref{eq:intcatch} this term is suppressed by 
$s_{21}^3$. 

Let us consider the condition of coherence restoration, 
which in the case of small density
is given by \Eqref{eq:cond12small}.  In the first approximation 
the condition is reduced to 
equality, $L_M \simeq L_C$. According to 
\Eqref{eq:lmantle} and (\ref{eq:lcore}) 
this condition is satisfied 
for the  nadir angle $\eta$ given by  
\be
\cos \eta \simeq \frac{3}{2\sqrt{2}} \sqrt{1 - \frac{R_C^2}{R_E^2}} 
= \frac{3}{\sqrt{2}} \cos \eta_c \simeq 0.89 \;,
\label{eq:etacat}
\ee
where $R_C$ and  $R_E$ are the radii of the core and the Earth; 
$\eta_c$ is the nadir angle of the trajectory which touches the core,  
$\cos \eta_c = 0.836$. The value $\cos\eta = 0.89$ gives $L_m = 3760\km$.

2) The term with phase $(2 \phi_1-\phi_2)$  in  \Eqref{eq:intcatch}
originates from  the interference of 
the 3rd and 6th terms of  \Eqref{eq:PcohEarth3}.
In turn these terms  are due to  the chains of transitions
\be
\nu_{1m}^{(1)} \rightarrow \nu_{2m}^{(2)} \rightarrow \nu_{1m}^{(1)}
\quad,\quad
\nu_{2m}^{(1)} \rightarrow \nu_{1m}^{(2)} \rightarrow \nu_{2m}^{(1)} \;.
\label{eq:tans2}
\ee   
This interference is a genuine 3-layer effect: 
In the first channel of (\ref{eq:tans2}) the WP is faster 
in the first mantle layer, then slower in the core,  
then again faster in the second mantle layer. 
In the second channel of (\ref{eq:tans2}) inversely: the WP moves first
more slowly, then faster, then again more slowly. 
This corresponds to the change of subscript indices of neutrino states in (\ref{eq:tans2}).  
So, the order of WP  in configuration space changes twice 
(the WP  of the first channel arrives first at the core, in the core the second WP 
overtakes the first one)
and catch-up occurs in the third layer.

The coherence restoration condition for this interference term is 
$2L_M \simeq L_C$,  which can be realized according 
to (\ref{eq:lmantle}) and (\ref{eq:lcore}) for 
\be
\cos \eta \simeq \frac{2}{\sqrt{3}} \sqrt{1 - \frac{R_c^2}{R_E^2}}
= \frac{2}{\sqrt{3}} \cos \eta_c \simeq 0.965 \;.
\label{eq:over2}
\ee
In this case $L_C = 3100\km$.
Unfortunately,  the amplitude of this 
interference term is suppressed very strongly by $s_{21}^4$. 

Equations (\ref{eq:etacat}) and (\ref{eq:over2})  are conditions for complete overlap of WP; 
partial overlap of WP and partial catch-up can be realized in a wider region of 
nadir angles and neutrino energies.   

3) The first term in \Eqref{eq:intcatch} does not contain phases, but we keep
it since it also shows the catch-up effect. The effect, however,
does not depend on energy and therefore it  
does not change by averaging over the energy and is present also
in the incoherent case.

There are two contribution to this term: interference of the
2nd and 5th terms in \Eqref{eq:PcohEarth3}, which both have the same phase
$\phi_1$, and interference of the 4th and 7th terms with common phase $\phi_1 + \phi_2$.
In the case of 2--5  interference one WP 
moves more slowly in the 3rd layer, whereas the other WP moves
more slowly in the 1st layer
(they move with the same high speed in the core).  So the second WP 
catches up with the first one  in the third layer.
The same is the case for 4--7 interference
with the only difference that in the core both WP move with the same
small speed. Here the catch-up does not depend on energy (if we neglect
the energy
dependence of the depth of interference). The derivatives of the
phases are the same for both channels,  and therefore
there is no averaging over the energy and the catch-up is complete.

This term is proportional to $s_{21}^2$ and thus less suppressed than the others.
It would acquire a phase if the 3rd layer was different from the first one.
However, in this case the coherence condition would be satisfied when
$L_3 \simeq L_1$. 

The amplitudes of the catch-up effect can be compared with
the main term in \Eqref{eq:PcohEarth3}, $(c_{10} c_{21} c_1)^2$, and
with the depth of the main oscillatory term (without loss of coherence). 
The latter  appears due to interference of the first term in \Eqref{eq:PcohEarth3}
with the 5th, 7th and 8th terms, which contain 
$s_{21}$ and $s_{10}$ to the first power. The depths equal 
$0.5 \sin 2\theta_{10}  \sin 2\theta_3$, 
$0.5 \sin 2\theta_{21}  \sin 2\theta_3 \simeq 0.08$
for $E = 15\MeV$.

The Earth matter effect and in particular the catch-up effect are further
suppressed because of the presence of the $\nu_2$ component in
the arriving supernova neutrino flux. 
For the $\nu_2 \to \nu_e$ transition the probability
can be obtained from the previous result by the substitutions
$c_{10} \rightarrow - s_{10}$,  $s_{10} \rightarrow c_{10}$.
As a result, all the interference terms \Eqref{eq:intcatch}
change sign, so the observable effect  will be proportional to 
the difference of the $\nu_1$  and $\nu_2$ fluxes,
$F(\nu_1) - F(\nu_2)$. In turn, this difference is determined by the
dynamics of flavor transitions in the supernova. An additional  suppression
can be estimated by a factor  $0.1 - 0.3$.

Taking $E=15\MeV$,  $\rho=5\g\cm^{-3}$ and $\rho=11\g\cm^{-3}$
for the densities of the Earth's mantle and core, respectively, we have
$\theta_1\simeq36^\circ$ and $\theta_2\simeq39^\circ$.  Consequently,
$\theta_{10}\simeq2.0^\circ$ and $\theta_{21}\simeq2.6^\circ$.
Approximating the energy dependence of combination of mixing parameters 
by power laws,
\Eqref{eq:intcatch} becomes roughly
\begin{align}
	P_\text{coh}^\text{int}(\nu_e) &\simeq
	1.4 \cdot 10^{-4} \left(\frac{E}{15\MeV}\right)^3 -
	9 \cdot 10^{-5}  \left(\frac{E}{15\MeV}\right)^3
	\cos(\phi_1-\phi_2) + {}
\nonumber\\
	&+ 1.5 \cdot 10^{-7}  \left(\frac{E}{15\MeV}\right)^5
	\cos(2\phi_1-\phi_2) \;.
\end{align}
The approximation is only accurate up to about $30\%$ for $5\MeV<E<80\MeV$,
but it serves to show that the oscillation depth inside the Earth is   small.
Consequently, the catch-up effect is too small (at the level $0.1\%$ at most) 
to be observable with
existing detectors, but it may become relevant in 
the future, especially if a supernova explodes close to us, producing
a huge number of events.

One last comment: as we have established, the separation and spread of
the WP have sizes of tens of meters. Such a state passes through the detector 
during about $t_\text{state} \sim 10^{-7}\,\text{sec}$.
So in principle, present technology allows to 
study parts of the packet. The problem is that the time of emission is not known 
and even the shortest features of the burst are about a few msec, which is much bigger than 
$t_\text{state}$.

\section{Conclusions}
%%%%%%%%%%%%%%%%%%%%%%%%%%%%%%%%%%%%%%%%%%%%%%%%%%%%%%%%%%%%%%%

Supernova neutrinos 
have several unique features: their wave packets (WP) are very short 
and coherence is lost very quickly during the propagation.
They propagate for a long time, so the spread of 
individual WP can reach macroscopic sizes,
up to hundreds of meters for a galactic supernova. 
Neutrino mass eigenstates arriving at the surface of the Earth split 
and oscillate again inside the Earth.

\begin{enumerate}
\item We have recalculated the size of the WP of supernova neutrinos,
finding a uniquely short length of about $10^{-11}\cm$, which
corresponds to an energy spread $\sigma_E \sim 1\MeV$ -- not much smaller than the
neutrino energy itself.
$\sigma_E$ does not depend on neutrino energy and is
approximately the same  for all neutrino species 
in all phases of the supernova.

\item The coherence length is smaller than $100\km$ for the 1-3 mass splitting
and of order $1000\km$ for the 1-2 mass splitting. The separation of the WP
arriving at the Earth from a supernova in the galactic center can be as 
large as $40\m$ for the 1-2 mass splitting and $E = 15\MeV$.

\item Each wave packet spreads due to the presence of different 
energy components in it. The spread is proportional to 
the neutrino mass squared and can reach a macroscopic 
size. Usually the separation of the packets 
is bigger that the spread. 

An exceptional situation is realized for neutrinos oscillating in the
Earth.  Here oscillations occur due to the interference of components
originating from the split of the WP of a mass eigenstate at the surface
of the Earth.  In this case the spread of the WP is much larger than
their relative shift inside the Earth, so they continue to overlap.  We
have showed, however, that this does not change the coherence condition,
which is determined by the original size of the WP without spread.

\item We have explored the coherence condition for supernova neutrinos 
oscillating in the Earth.
The coherence length turns out to be comparable with the sizes of the
mantle and the core.  Thus, for low energies 
($E < 30\MeV$) coherence is completely lost for most nadir angles. 
For a large range of energies and nadir angles the loss of coherence is partial. Only for high energies and 
shallow trajectories decoherence can be neglected.

\item We have studied oscillations in a multi-layer medium 
characterized by the adiabatic change of density 
within layers and sudden jumps of density between layers. 
This has applications for neutrinos crossing the shock waves in a
supernova and for neutrinos propagating inside the Earth along 
core-crossing trajectories.
 
\item A splitting of eigenstates occurs at each density jump, so for
two-neutrino mixing $2^n$ 
components are formed after $n$ crossings. This multiplication has no meaning if 
the shift of the WP within each layer can be neglected. 
In this case the problem is reduced to the two-neutrino problem. 
However, if the shift and therefore decoherence are 
substantial, the multiple splitting has physical sense
and can lead to new phenomena.
  
In particular, it leads to the new interesting phenomenon
of the partial restoration of coherence due to a
``catch-up effect''. 
In other words, in a multi-layer medium 
coherence can be partially restored and the
coherence length can be increased beyond the usual estimate. 
In the simplest realization, for two layers, this happens 
if a component of the WP that
travels faster through layer $1$ arrives at a detector at the same time
as a component that was slower in layer $1$ but faster in layer $2$.
The described effect yields
corrections at the percent level or below for supernova neutrinos
oscillating in Earth matter, but in principle it can be observed if a
very high-statistics signal from a close supernova is detected.

\item We have studied decoherence
in parallel in configuration and momentum space, checking the
equivalence between both representations.  Although at first sight the
catch-up effect seems to depend on the size of the WP,
this is not the case in the
examples we have considered.  We have explicitly shown how this 
size-independence is due to the
restoration of overlap of WP,
confirming the general (abstract) results of \cite{Kiers:1995zj}
and \cite{Stodolsky:1998tc}.
In a sense the catch-up effect is a non-trivial effect of the averaging
of oscillation probabilities over energy.  

We have verified that in all cases we have studied there is an equivalence
between configuration and momentum space.  That is, we can choose to do
all calculations either with WP that separate or by suitably
averaging over energy.  
The observable oscillatory picture is determined by the initial 
energy spectrum, the energy resolution of the detector, and by the phase
acquired between source and detector
as a function of energy, unless time tagging is arranged. 
\end{enumerate}

\subsection*{Acknowledgements}
We would like to thank Evgeny Akhmedov, Yasaman Farzan, Carlo Giunti, Alessandro Mirizzi,
Georg Raffelt, G\"unter Sigl, Irene Tamborra, Ricard Tomas Bayo, and Mariam Tortola Baixauli
for helpful discussions.
We acknowledge support from the European Union FP7 ITN
In$\nu$isibles (Marie Curie Actions, PITN-GA-2011-289442) and from the
Max Planck fellowship M.FW.A.KERN0001.

\frenchspacing
\bibliography{Neutrinos}
\bibliographystyle{utphys}

\end{document}